\documentclass[aps,prd,notitlepage,amssymb,amsmath,floatfix,nofootinbib,superscriptaddress]{revtex4-1}

\usepackage{epsfig}
\usepackage{bm}
\usepackage{amssymb}
\usepackage{amsmath}
\usepackage{color}
\usepackage{slashed}
\usepackage{subfigure}
\usepackage[colorlinks,linkcolor=blue,anchorcolor=black,citecolor=blue]{hyperref}
\newcommand{\bs}{\boldsymbol}
\allowdisplaybreaks[4]

\begin{document}

\title{Transverse polarization of Lambda hyperons in hadronic collisions}

\author{Ying Gao}
\affiliation{School of Physics and Technology, University of Jinan, Jinan, Shandong 250022, China}

\author{Kai-Bao Chen}
\email{chenkaibao19@sdjzu.edu.cn}
\affiliation{School of Science, Shandong Jianzhu University, Jinan, Shandong 250101, China}

\author{Yu-Kun Song}
\email{sps\_songyk@ujn.edu.cn}
\affiliation{School of Physics and Technology, University of Jinan, Jinan, Shandong 250022, China}

\author{Shu-Yi Wei}
\email{shuyi@sdu.edu.cn}
\affiliation{Institute of Frontier and Interdisciplinary Science, Key Laboratory of Particle Physics and Particle Irradiation (MOE), Shandong University, Qingdao, Shandong 266237, China}

\begin{abstract}
The transverse polarization of $\Lambda$ hyperon within reconstructed jets in hadronic collisions offers a complementary platform to probe the polarized fragmentation function $D_{1T}^\perp$. We illustrate that by performing a global analysis of the transverse polarization of $\Lambda$ hyperons produced in different kinematic regions and in different hadronic collisions, such as $pp$, $p\bar p$, $pA$, and $\gamma A$ collisions, we can pin down the flavor dependence of $D_{1T}^\perp$ which has been poorly constrained. Besides the single inclusive jet production, the $\gamma/Z^0$-boson associated jet production supplements with more capability in removing ambiguities in the flavor dependence of $D_{1T}^\perp$.
\end{abstract}

\maketitle

\section{Introduction}

Fragmentation functions (FFs) are nonperturbative quantities in calculating hadron production cross section in high energy reactions~\cite{Collins:1981uw, Metz:2016swz, Chen:2023kqw, Boussarie:2023izj}, which are still not eligible for first-principle calculations in QCD. Usually, a global analysis of experimental data can provide vital information on FFs, which can then be used in making predictions for future experiments. On one hand, there is quite a lot of experimental data on the single inclusive hadron production in various high-energy collisions, which is sensitive to the collinear unpolarized FF $D_{1} (z)$. Therefore, quite a few studies~\cite{Binnewies:1994ju, Kretzer:2000yf, Bourhis:2000gs, deFlorian:2007aj, deFlorian:2007ekg, Albino:2008fy, deFlorian:2014xna, deFlorian:2017lwf, Borsa:2021ran, Borsa:2022vvp, Borsa:2023zxk} have extracted it with a global analysis. On the other hand, the transverse-momentum-dependent fragmentation functions (TMD FFs), especially the spin-dependent ones, are largely unknown due to limited data. Nonetheless, it has still attracted a lot of attention recently. Many phenomenological studies devoted to this topic~\cite{Anselmino:2007fs, Anselmino:2013vqa, Anselmino:2013lza, Anselmino:2015sxa, Kang:2015msa, Anselmino:2015fty, Bacchetta:2001di, Metz:2002iz, Gamberg:2003eg, Amrath:2005gv, Bacchetta:2007wc, Matevosyan:2012ga, Lu:2015wja, Yang:2016mxl, Yang:2017cwi, Wang:2018wqo, Xie:2022lra}. Furthermore, the QCD evolution~\cite{Aybat:2011ge, Echevarria:2014rua} for the spin-dependent fragmentation functions has also been established at the leading twist, which paves the way for a global analysis of experimental data at different energy scales in the future. For a recent review of the general progress, see e.g.~\cite{Metz:2016swz, Chen:2023kqw, Boussarie:2023izj}.

The polarized FF $D_{1T}^\perp$, which is the partner in FFs of the famous Sivers function, induces the transverse polarization along the normal direction of the hadron production plane, i.e., along the $\bm{n}\sim \bm{k}_j\times \bm{p}_h$ direction with $\bm{k}_j$ and $\bm{p}_h$ being the three-momenta of the fragmenting parton and the produced hadron. Belle collaboration has measured the transverse polarization of $\Lambda$ in inclusive and semi-inclusive $e^+e^-$-annihilations~\cite{Belle:2018ttu}, which has been immediately followed by several phenomenological studies~\cite{Anselmino:2019cqd, DAlesio:2020wjq, Callos:2020qtu, Chen:2021hdn, Li:2020oto, Matevosyan:2018jht, DAlesio:2021dcx, Gamberg:2018fwy, Gamberg:2021iat, Yang:2021zgy, Li:2021txj}. A major issue emerges \cite{Chen:2021hdn} in these studies: is the isospin symmetry of the polarized FF violated? The answer to this question is crucially important in understanding the fundamental properties of strong interaction. However, as confirmed in Refs.~\cite{DAlesio:2022brl,DAlesio:2023ozw}, current experimental data from Belle alone cannot settle this issue. Therefore, it has been proposed that the future measurement of Lambda transverse polarization at the future electron-ion colliders (EIC) can ultimately test the isospin symmetry~\cite{Chen:2021zrr}. Similar studies have also been presented in Refs.~\cite{Kang:2021kpt, DAlesio:2023ozw}. With the EIC experiment still at the horizon, we look for opportunities in experimental facilities that are currently available.

The concept of {\it hadron within jet} has been proposed~\cite{Yuan:2007nd, Bain:2016rrv, Kang:2016ehg, Kang:2017glf, Kang:2020xyq, Kang:2023elg}, which makes it possible to study TMD FFs even in hadronic collisions. Recent works~\cite{Kang:2020xyq, DAlesio:2024ope} have also constructed a framework to study $D_{1T}^\perp$ in $pp$ collisions. The relation between jet FFs and the conventional TMD parton FFs has been established. Preliminary results from the STAR collaboration have also been published \cite{Gao:2024dxl}, which indicates the potential of studying spin effects in unpolarized hadronic collisions. Furthermore, hadronic collisions also provide an ideal platform to investigate the gluon FFs. As elaborated in Refs.~\cite{Zhang:2023ugf, Li:2023qgj}, the longitudinal spin transfer of the gluon can be well constrained by measuring the longitudinal polarization correlation of two final state hadrons. 

In this work, we investigate the Lambda transverse polarization in different hadronic collisions and demonstrate its potential of extracting the flavor dependence of the polarized FF. We benefit from the hadron within jet framework~\cite{Yuan:2007nd, Bain:2016rrv, Kang:2017glf, Kang:2020xyq, Kang:2023elg, DAlesio:2024ope} and focus mainly on the phenomenological side. Therefore, we work at leading order (LO) in perturbative QCD and identify the TMD jet FFs with the TMD parton FFs. We perform a comprehensive calculation for the Lambda transverse polarizations in $pp/p\bar p/pA/\gamma A$ collisions. We demonstrate that the flavor components are quite different in different kinematic regions and in different hadronic collisions. Therefore, the current hadron colliders, such as RHIC, LHC and Tevatron, can significantly improve our understanding on the polarized FF. We also present our numerical results for the Lambda transverse polarization within $\gamma/Z^0$-boson associated jets. The flavor component is different from that in the single inclusive jet production. Therefore, it provides a complementary channel to investigate the flavor dependence.

The jet FFs are in principle different from the parton FFs. However, as demonstrated in \cite{Kang:2016ehg,Kang:2017glf,Kang:2023elg}, they can be matched to the latter. The matching coefficient is perturbatively calculable. Therefore, an extended universality has been established between parton FFs and jet FFs. At the LO accuracy, they are identical. Since the aim of this paper is to demonstrate the potential of hadron colliders for disentangling the flavor dependence of $D_{1T}^\perp$, we perform a LO calculation and leave a more accurate NLO global analysis for a future study when more experimental data are available. The LO calculation contains large uncertainties. For instance, the LO cross section strongly depends on the the choice of factorization and renormalization scales. However, in the context of flavor components, the scale dependence becomes negligible. A more precise next-to-leading order (NLO) calculation only provides a small correction and thus does not affect the qualitative conclusion. We present a comparison between LO and NLO results in Appendix B. 

The rest of the paper is organized as follows. In Sec. II, we present the general formulas to calculate the jet production ratios and Lambda transverse polarization in hadronic collisions. In Sec. III, we present our numerical results for the production ratios and polarization in various reactions, and extensively explore the potential for flavor separation of $D_{1T}^\perp$. A summary is given in Sec. IV.

\section{General formulas}

In this section, we first lay out the formulas for calculating hadron-within-jet cross section in hadronic collisions. Then, we present the Lambda transverse polarization with a Gaussian ansatz.

\subsection{Differential cross sections}

Since our goal is to investigate the potential of extracting the flavor dependence of the polarized FF in hadronic collisions, we first present the differential cross section of the single inclusive jet production with a given flavor in $pp$ collisions. Although the flavor of a reconstructed jet in experiments is not well defined, we still employ this phrase since we have limited ourselves at the LO approximation. A jet is simply a parton at LO. The flavor dependent cross section is given by
\begin{align}
  \frac{d\sigma_{p+p\to {j} + X}}{dyd^2\bs{k}_T} = \sum_{abc} \int d y_2  x_1 f_a (x_1,\mu_f) x_2 f_b(x_2,\mu_f)  \frac{1}{\pi} \frac{d\hat\sigma_{a+b\to j+c}}{d\hat t},
\end{align}
where $f_{a/b}$ is the parton distribution function with $x_i$ the momentum fraction of the hadron carried by the parton and $\mu_f$ the factorization scale, $y$ and $\bs{k}_T$ denote the rapidity and transverse momentum of parton $j$ which is also identified as the jet measured in experiments, and $y_2$ denotes the rapidity of unmeasured final state parton $c$. The momentum fractions $x_1$ and $x_2$ can be calculated with $x_1 = {k_T} (e^{y} + e^{y_2}) /\sqrt{s}$ and $x_2 = {k_T} (e^{-y} + e^{-y_2}) /\sqrt{s}$ for massless partons. $d\hat\sigma_{a+b\to j+c}/d\hat t$ denotes the partonic cross section. To compare with the single inclusive jet cross section measured in the experiments, we have to sum over all possible ``jet flavors'' since current experimental instruments cannot distinguish jet flavors. 

To quantify the flavor component of final state jets, we define the production ratio of a given jet flavor as 
\begin{align}
R_j(y,k_T) \equiv \frac{d\sigma_{p+p\to {j}+X}/dyd^2\bs{k}_T}{\sum_{i}d\sigma_{p+p\to {i}+X}/dyd^2\bs{k}_T}.
\end{align}

The cross section of hadron within jet production at the LO can be obtained by convoluting the above jet cross section with the corresponding jet FF~\cite{Yuan:2007nd, Kang:2017btw}. We obtain 
\begin{align}
  \frac{d\sigma_{p+p\to j(\to \Lambda)+X}}{dyd^2\bs{k}_Tdzd^2\bs{p}_{\Lambda T}}=\sum_j \frac{d\sigma_{p+p\to j+ X}}{dyd^2\bs{k}_T}\left(D_{1j}^\Lambda(z,\bs{p}_{\Lambda T})+\frac{\epsilon_\perp^{\rho\sigma}p_{\Lambda T\rho}S_{\Lambda T\sigma}}{zM_\Lambda}D_{1T,j}^{\perp\Lambda}(z,\bs{p}_{\Lambda T})\right),
\label{eq:cs-hainje}
\end{align}
where $z$ is the momentum fraction of the parton carried by the final state hadron and $\bm{p}_{\Lambda T}$ is the transverse momentum of $\Lambda$ with respect to the jet momentum. Here $M_\Lambda$ is the Lambda mass and $S_{\Lambda T}$ is the transverse spin vector of $\Lambda$.  $D_{1j}^\Lambda$ is the unpolarized FF, and $D_{1T,j}^{\perp\Lambda}$ is the polarized FF.

\subsection{Transverse polarization}

The transverse polarization of final state $\Lambda$ hyperons along the $\bs{n} \equiv \frac{\bs{k}_j\times \bs{p}_{\Lambda}}{|\bs{k}_j\times \bs{p}_{\Lambda}|}$ direction is given by
\begin{align}
  &P_\Lambda(y,k_T,z,p_{\Lambda T}) \equiv \left(\frac{d\sigma_{p+p\to j(\to\Lambda^\uparrow)+ X}}{dyd^2\bs{k}_Tdzd^2\bs{p}_{\Lambda T}}-\frac{d\sigma_{p+p\to j(\to\Lambda^\downarrow)+ X}}{dyd^2\bs{k}_Tdzd^2\bs{p}_{\Lambda T}}\right)\Bigg/\left(\frac{d\sigma_{p+p\to j(\to\Lambda^\uparrow )+ X}}{dyd^2\bs{k}_Tdzd^2\bs{p}_{\Lambda T}}+\frac{d\sigma_{p+p\to j(\to\Lambda^\downarrow )+ X}}{dyd^2\bs{k}_Tdzd^2\bs{p}_{\Lambda T}}\right).
\label{eq:pola-def}
\end{align}
Here $\Lambda^{\uparrow}$ indicates $\bm{S}_{\Lambda T} = \bm{n}$, and $\Lambda^{\downarrow}$ means $\bm{S}_{\Lambda T} = -\bm{n}$. 

Inserting Eq.~(\ref{eq:cs-hainje}) into Eq.~(\ref{eq:pola-def}), it is straightforward to obtain
\begin{align}
P_\Lambda(y,k_T,z,p_{\Lambda T})
&= \sum_j \frac{d\sigma_{p+p\to j+ X}}{dy d^2\bs{k}_T} \frac{p_{\Lambda T}}{zM_\Lambda} D_{1T,j}^{\perp\Lambda}(z,\bs{p}_{\Lambda T})/\sum_j \frac{d\sigma_{p+p\to j+ X}}{dyd^2\bs{k}_T} D_{1j}^\Lambda(z,\bs{p}_{\Lambda T})
\nonumber\\
&=\sum_j R_j(y,k_T) \frac{p_{\Lambda T}}{zM_\Lambda}D_{1T,j}^{\perp\Lambda}(z,\bs{p}_{\Lambda T})/\sum_j R_j(y,k_T) D_{1j}^\Lambda(z,\bs{p}_{\Lambda T}).
\end{align}
Since this paper mainly focuses on the flavor dependence of the polarized FF $D_{1T,j}^{\perp\Lambda}$, we neglect the QCD evolution effect for simplicity and take Gaussian ansatz for $\bs{p}_{\Lambda T}$ distribution, namely
\begin{align}
D_{1j}^\Lambda(z,\bs{p}_{\Lambda T})&=D_{1j}^\Lambda(z)\frac{1}{\pi \Delta_\Lambda^2}e^{-\bs{p}_{\Lambda T}^2/\Delta_\Lambda^2},\\
D_{1T,j}^{\perp\Lambda}(z,\bs{p}_{\Lambda T})&=D_{1T,j}^{\perp \Lambda}(z)\frac{1}{\pi \Delta_\Lambda^2}e^{-\bs{p}_{\Lambda T}^2/\Delta_\Lambda^2}.
\end{align}
Here, $\Delta_\Lambda$ is the Gaussian width which, as an approximation, is chosen the same between the unpolarized and polarized FFs. With the Gaussian ansatz and taking average over the magnitude of $k_T$, the final expression for the Lambda polarization can be simplified to
\begin{align}
P_\Lambda (y,k_T,z)=\frac{\sqrt{\pi}\Delta_\Lambda}{2zM_\Lambda}\frac{\sum_j R_j(y,k_T) D_{1Tj}^{\perp\Lambda}(z)}{\sum_j R_j(y,k_T) D_{1j}^\Lambda(z)}.
\end{align}

In the numerical calculation, we employ the DSV parametrization~\cite{deFlorian:1997zj} to provide the unpolarized Lambda FF $D_{1j}^\Lambda$ and mainly employ the isospin symmetric CLPSW parametrization~\cite{Chen:2021hdn} for the polarized one $D_{1T,j}^\perp$. For comparison, we have also used the isospin violated DMZ~\cite{DAlesio:2020wjq} and CKT~\cite{Callos:2020qtu} parametrizations in some calculations. 

Furthermore, in the hybrid factorization framework, the factorization scale of the initial state collinear parton distribution functions (PDFs) is usually chosen to be $k_T$ with $k_T$ being the transverse momentum of the jet in the lab frame. However, that of the final state TMD jet FFs is usually chosen to be $k_T R$ with $R$ being the jet cone size. In the numerical evaluation, the DGLAP evolution of collinear PDFs has been taken care of by the CT18 PDF \cite{Hou:2019efy}. 
The QCD evolution of the final state $D_{1T}^\perp$ has been extensively investigated in~\cite{Gamberg:2021iat,Kang:2010xv}. As discussed in Appendix C, this effect also does not change our conclusion. Therefore, we take the Gaussian ansatz for simplicity.

\section{Numerical results}

In this section, we present our numerical predictions for the Lambda polarization in different hadronic collisions, namely $pp$, $p\bar p$, $pA$, and photon-nucleus collisions. We make predictions for the single inclusive jet production process as well as the $\gamma/Z^0$-boson associated jet production process. The flavor components in different processes vary quite a lot. Therefore, a global analysis of all those processes can eventually lay out the flavor dependence of the polarized fragmentation function. 

\subsection{Single inclusive jet production in $pp$, $p\bar p$ and $pA$ collisions}

\begin{figure}[htb!]
  \centering
  \includegraphics[width=0.9\textwidth]{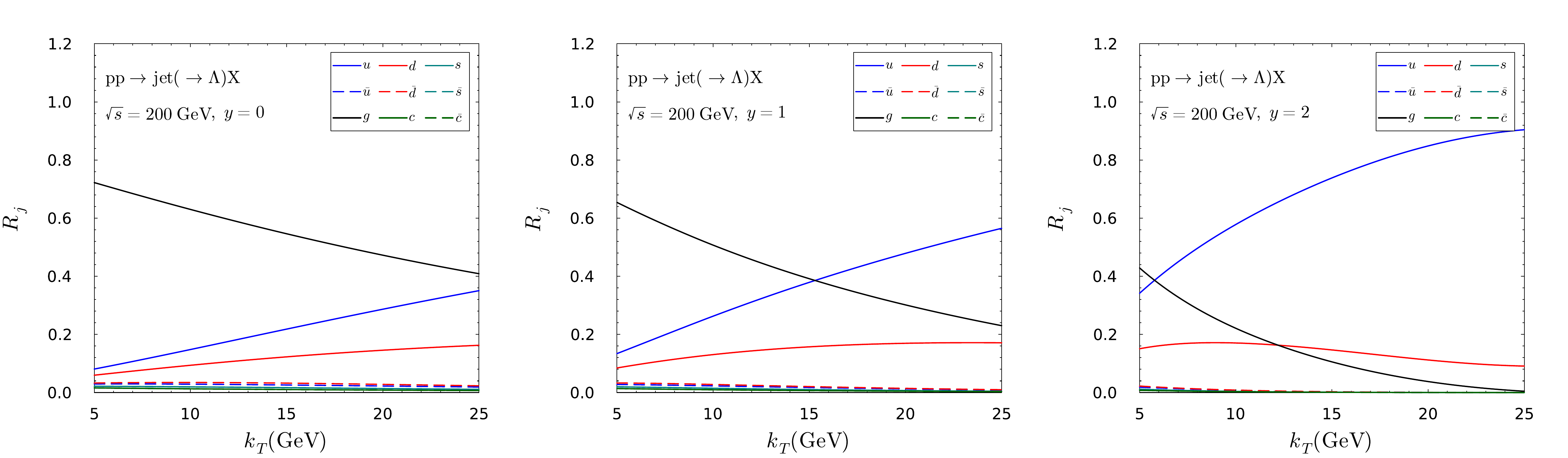}\\
  \includegraphics[width=0.9\textwidth]{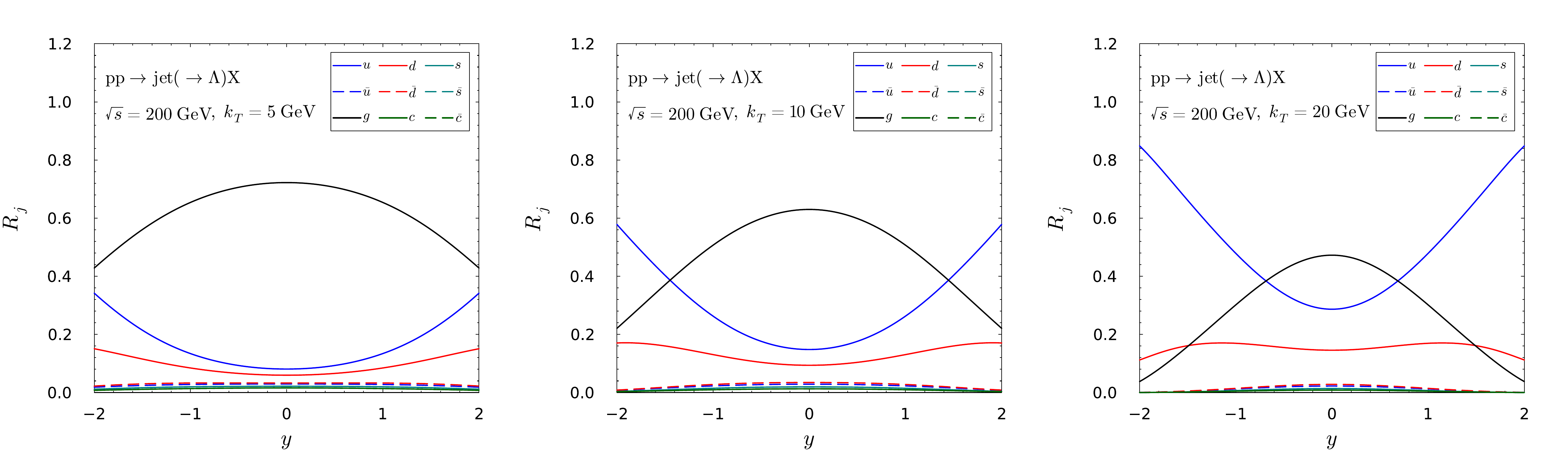}
  \caption{Production ratios of different flavor jets in $pp$ collisions at $\sqrt{s}=200$ GeV. Here and in the following figures we employed the CT18 PDFs~\cite{Hou:2019efy} in the numerical evaluation.}
  \label{fig:Rq_pp_200GeV}
\end{figure}

To demonstrate the potential of revealing the flavor dependence, we first calculate the production ratios of different flavor jets in $pp$ collisions at the RHIC energy. The numerical results at the LO accuracy are presented as a function of jet $k_T$ in the upper panel of Fig.~\ref{fig:Rq_pp_200GeV} and as a function of jet rapidity in the lower panel, where we employ CT18 PDFs~\cite{Hou:2019efy}. The qualitative feature is well-known: the valance quark contribution increases with increasing $k_T$ and increasing rapidity, while the gluon contribution decreases. The same feature is also exhibited at the LHC energies, as shown in Figs.~\ref{fig:Rq_pp_y0_LHC} and~\ref{fig:Rq_pp_pt0_LHC} in the appendix. Therefore, by comparing the Lambda transverse polarization at different rapidities and jet transverse momenta, we obtain the power to extract the polarized gluon FFs which has been poorly constrained in previous studies~\cite{DAlesio:2020wjq,Callos:2020qtu,Chen:2021hdn}.

\begin{figure}[htb!]
  \centering
  \includegraphics[width=0.9\textwidth]{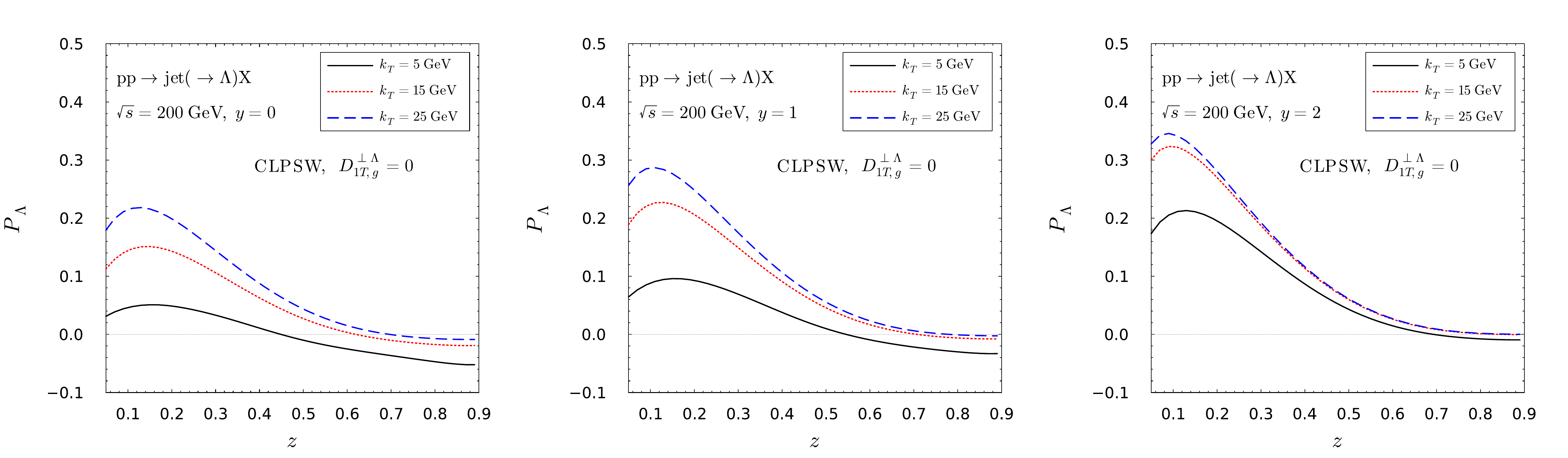}
  \includegraphics[width=0.9\textwidth]{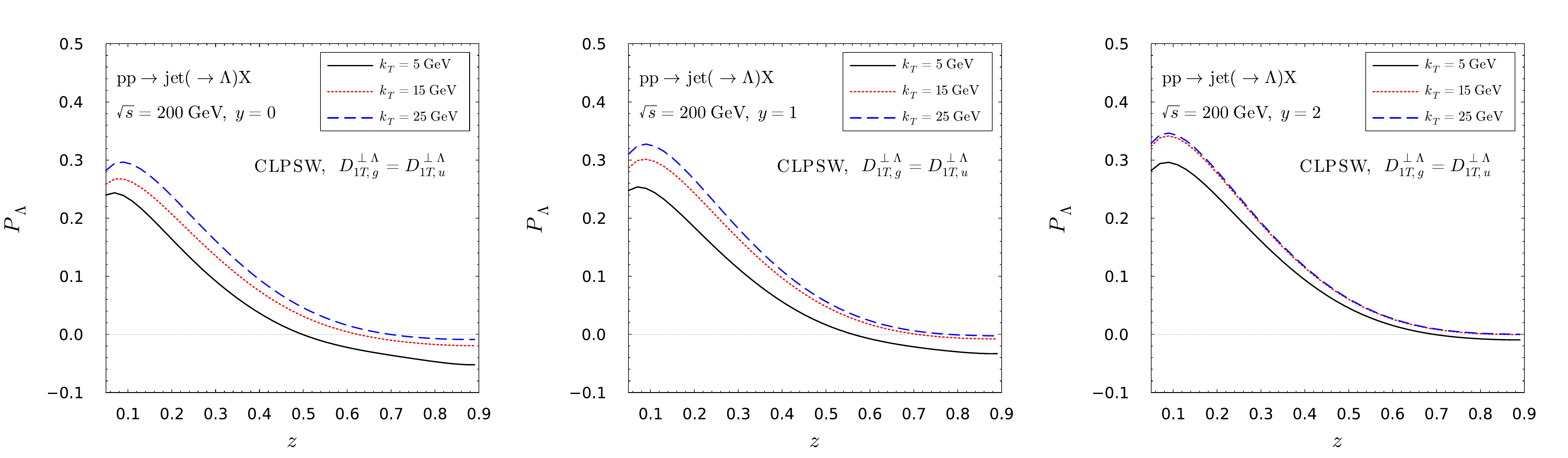}
  \caption{Transverse polarization of $\Lambda$ as functions of $z$ at different $y$ and $k_T$ in $pp$ collisions. The upper panel shows the numerical results assuming $D_{1T,g}^\perp = 0$, while the lower panel shows those assuming $D_{1T,g}^{\perp} = D_{1T,u}^\perp = D_{1T,d}^\perp$.}
  \label{fig:Polarization_pp_200GeV}
\end{figure}

For instance, the CLPSW parametrization~\cite{Chen:2021hdn} is proposed to describe the Lambda transverse polarization in $e^+e^-$ annihilation at LO. Therefore, the gluon fragmentation function is absent in the previous analysis. However, it is important in making predictions for the hadronic collisions, since they amount to a sizable contribution. In this study, we propose two extreme prescriptions. In the first prescription, we simply assume that the gluon does not contribute to the Lambda polarization by setting $D_{1T,g}^{\perp} = 0$. In the second prescription, we assume that $D_{1T,g}^\perp = D_{1T,u}^\perp = D_{1T,d}^\perp$. The numerical results are shown in Fig. \ref{fig:Polarization_pp_200GeV}. The upper panel shows the results using the first prescription, while the lower panel shows those in the second prescription. The real case should deviate from these two simple prescriptions and might be somewhere in between. However, the difference between two panels demonstrates this process is indeed sensitive to the gluon FF. Therefore, it is plausible to extract the gluon FF. More numerical predictions for the LHC energies are presented in the appendix. 

\begin{figure}[htb!]
  \centering
    \includegraphics[width=0.9\textwidth]{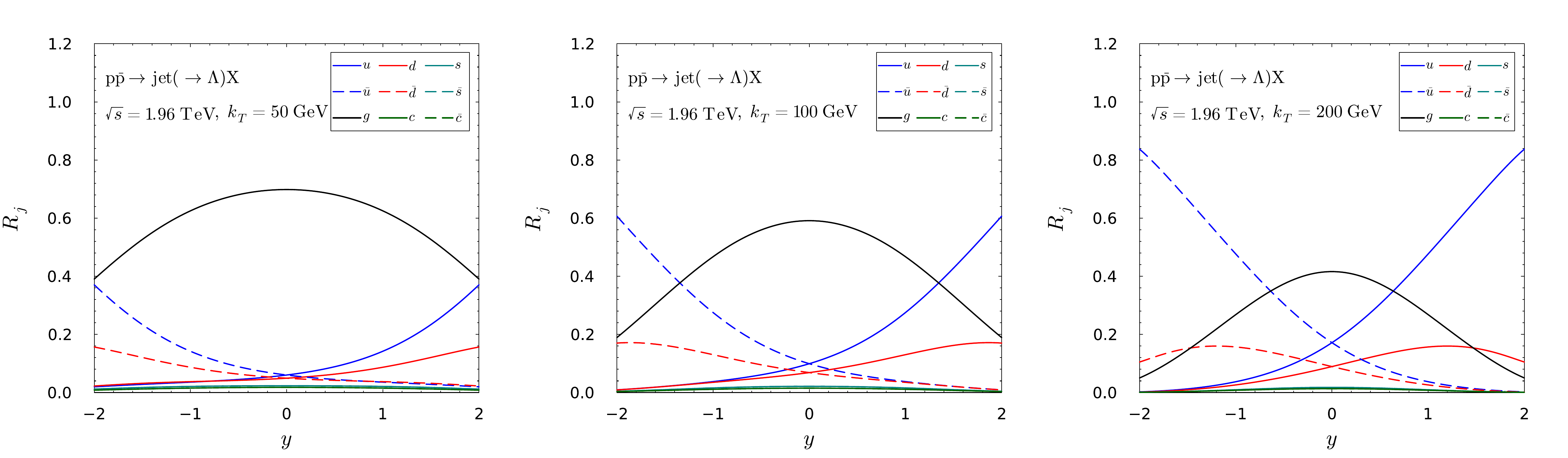}
  \caption{Production ratios of different flavor jets as functions of rapidity $y$ at fixed $k_T$ in $p\bar p$ collisions. We have employed the CT18~\cite{Hou:2019efy} PDFs in the numerical calculation.}
  \label{fig:Rq_ppbar}
\end{figure}

\begin{figure}[htb!]
  \centering
  \includegraphics[width=0.9\textwidth]{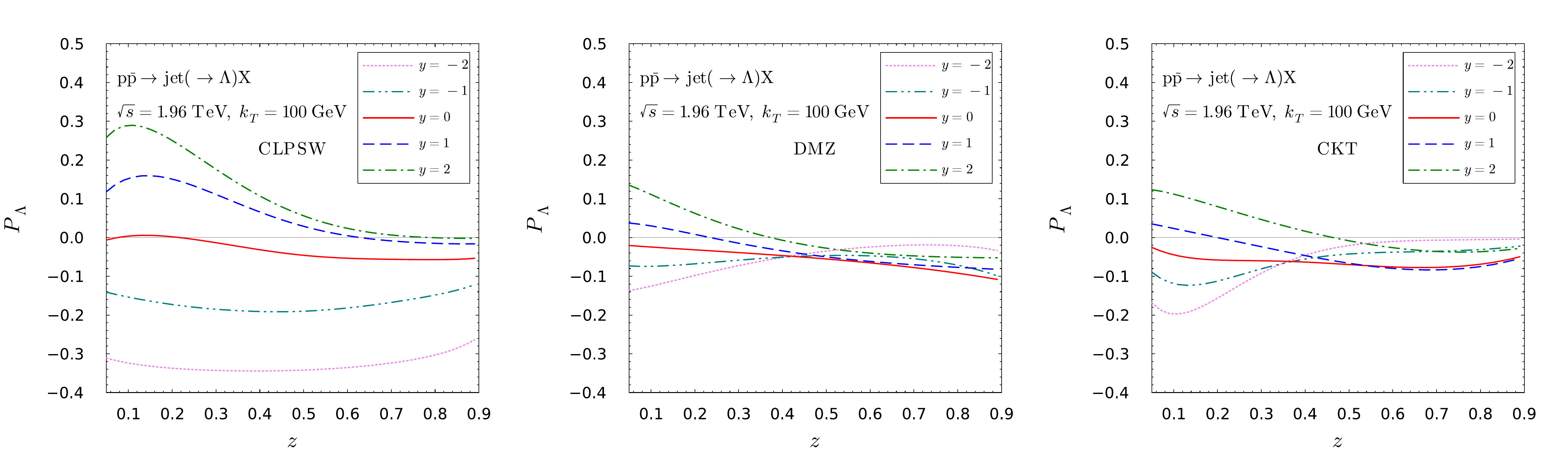}
  \caption{Transverse polarization of $\Lambda$ as functions of $z$ at different $y$ and $k_T$ in $p\bar p$ collisions.}
  \label{fig:Polarization_ppbar}
\end{figure}

The second process we investigate is the single inclusive jet production in $p\bar p$ collisions. Unlike the case in $pp$ collisions, the forward and backward directions are no longer symmetric in $p\bar p$ collisions. We show the production rate of different flavor jets as a function of rapidity in Fig.~\ref{fig:Rq_ppbar}. It is apparent that there are more $u$ quark jets in the forward rapidity (proton-going direction) and there are more $\bar u$ quark jets in the backward rapidity (antiproton-going direction). This forward-backward asymmetric enables us to examine the difference between quark and antiquark fragmentation functions. This point can be fully appreciated from the numerical results shown in Fig.~\ref{fig:Polarization_ppbar}. The CLPSW, DMZ and CKT parametrizations have parametrized the $u$ and $\bar u$ fragmentation differently, so that the magnitude of forward-backward asymmetry also varies. Therefore, including the Tevatron experiment in the global analysis can cast more light on the flavor dependence of $D_{1T}^\perp$.

\begin{figure}[htb!]
  \centering
    \includegraphics[width=0.27\textwidth]{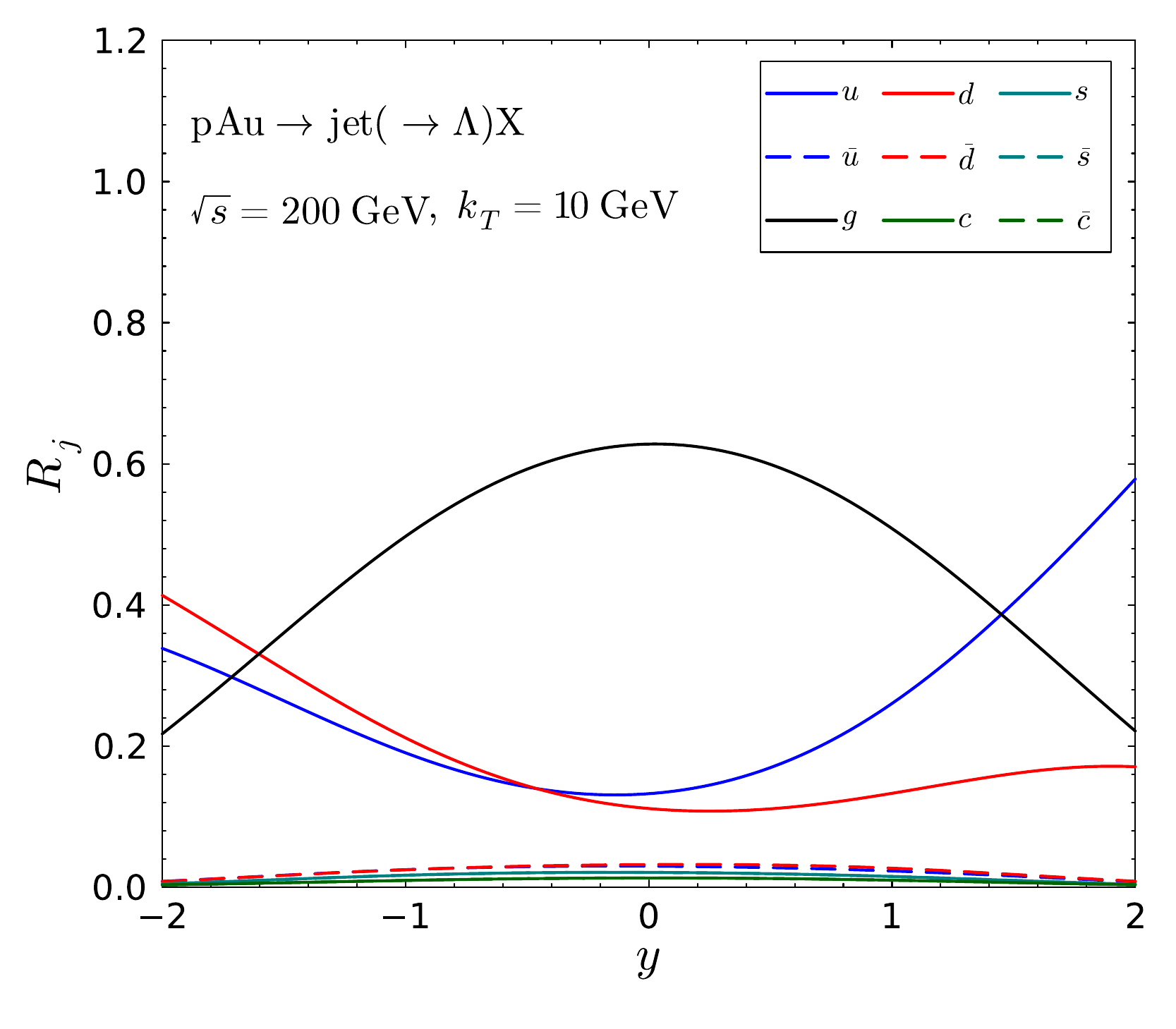}~~~
    \includegraphics[width=0.27\textwidth]{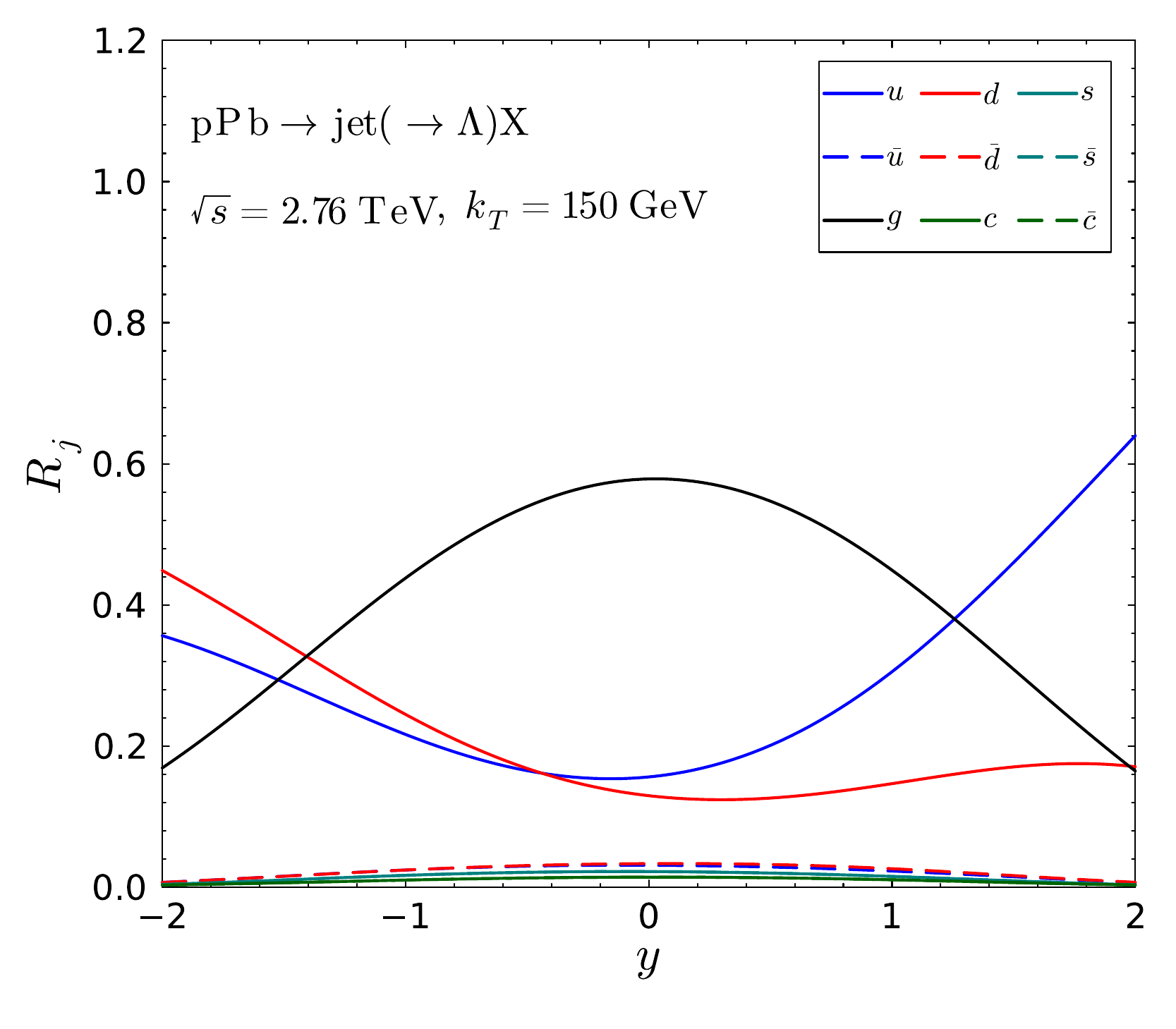}~~~
    \includegraphics[width=0.27\textwidth]{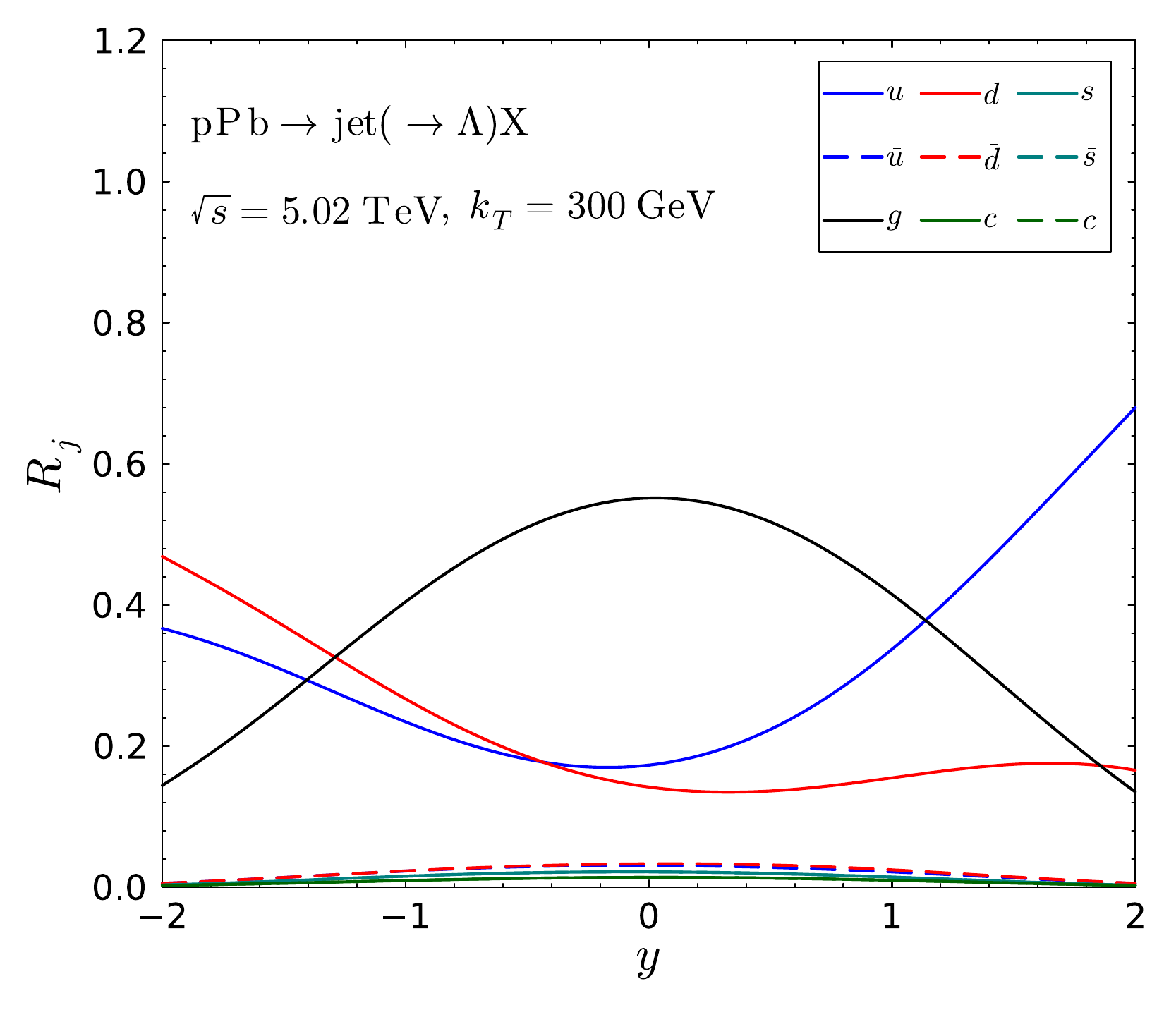}
  \caption{Production ratios of different flavor jets as functions of $y$ in $pA$ collisions. Here and in the following figures the nuclear modified parton distributions are given by the EPPS21 sets~\cite{Eskola:2021nhw}.}
  \label{fig:Rq_pA}
\end{figure}

The last process we study in this subsection is the single inclusive jet production in $pA$ collisions. Similar to the $p\bar p$ collisions, there is also a forward-backward asymmetry. However, in this process, we can study the difference between $u$ and $d$ quark fragmentation function. We present the production ratios of different flavor jets as a function of jet rapidity in Fig.~\ref{fig:Rq_pA}, which clearly shows that there are more $u$ quark jets in the proton-going direction and more $d$ quark jets in the nucleus-going direction. For nuclear modified PDFs we take EPPS21 sets~\cite{Eskola:2021nhw}.

\begin{figure}[htb!]
  \centering
  \includegraphics[width=0.9\textwidth]{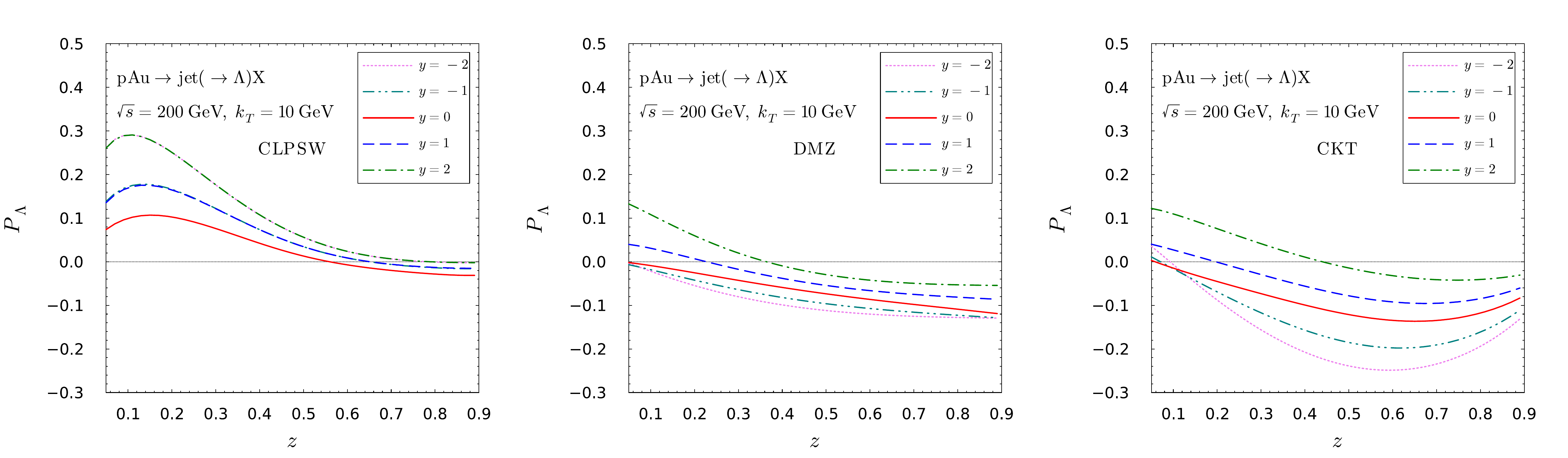}\\
  \includegraphics[width=0.9\textwidth]{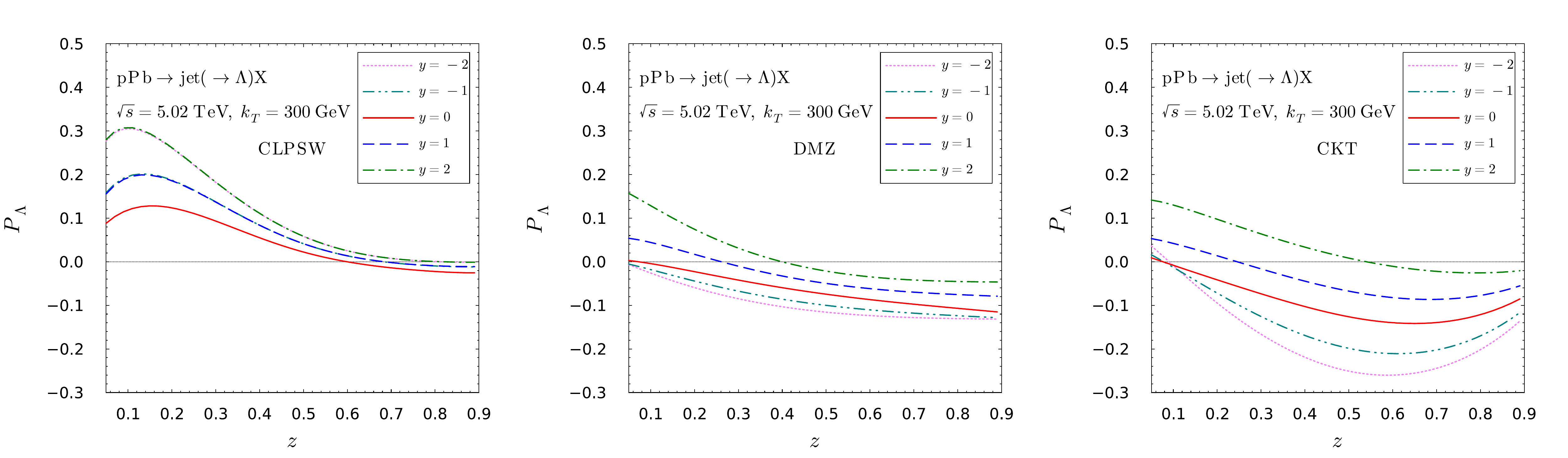}
  \caption{Transverse polarization of $\Lambda$ as functions of $z$ at different $y$ and $k_T$ in $pA$ collisions.}
  \label{fig:Polarization_pA}
\end{figure}

As underlined in the previous work~\cite{Chen:2021hdn}, the Lambda hyperon is an isospin singlet. The isospin symmetry requires $D_{1T,u}^{\perp} = D_{1T,d}^{\perp}$. This requirement has been respected in the CLPSW parametrization, while it has been severely violated in the DMZ and CKT parametrizations. The answer to whether or not the isospin symmetry is violated in the hadronization process is vitally important for understanding strong interaction at the non-perturbative regime. We thus present the numerical results for the Lambda polarization at different rapidities in $pA$ collisions in Fig.~\ref{fig:Polarization_pA}. Apparently, there is almost no difference between forward and backward rapidities in the isospin symmetric CLPSW parametrization. However, one can observe distinguish difference in the DMZ and CKT parametrizations. Therefore, future experimental measurements on this aspect can eventually clarify this issue.

To summarize, a global analysis of the Lambda transverse polarization in different kinematic regions in $pp$, $p\bar p$, and $pA$ collisions has great potential in pinning down the flavor dependence of the polarized fragmentation function $D_{1T}^\perp$.

\subsection{$\gamma/Z^0$-boson associated jet production}

Besides the single-inclusive jet production, another important channel deserving study is the $\gamma/Z^0$-boson associated jet production. Unlike the single-inclusive jet production dominated by strong interaction, this process is an electroweak process. The electroweak vertex modifies the production ratios of jets with different flavors. Therefore, the phenomenological study of this channel provides a complementary platform to investigate flavor dependence. 

To be more specific, the process we consider is $p + p \to {\rm jet} (\to \Lambda) + \gamma/Z^0 + X$. Here, we only consider the prompt production for the final state photon or $Z^0$-boson. At LO, there are two partonic channels that contribute, namely $q + g \to q + \gamma/Z^0$ and $q + \bar q \to g+\gamma/Z^0$.  Therefore, the gluon contribution is strongly suppressed at low jet $k_T$ and in the central rapidity. 

\begin{figure}[htb!]
  \centering
  \includegraphics[width=0.9\textwidth]{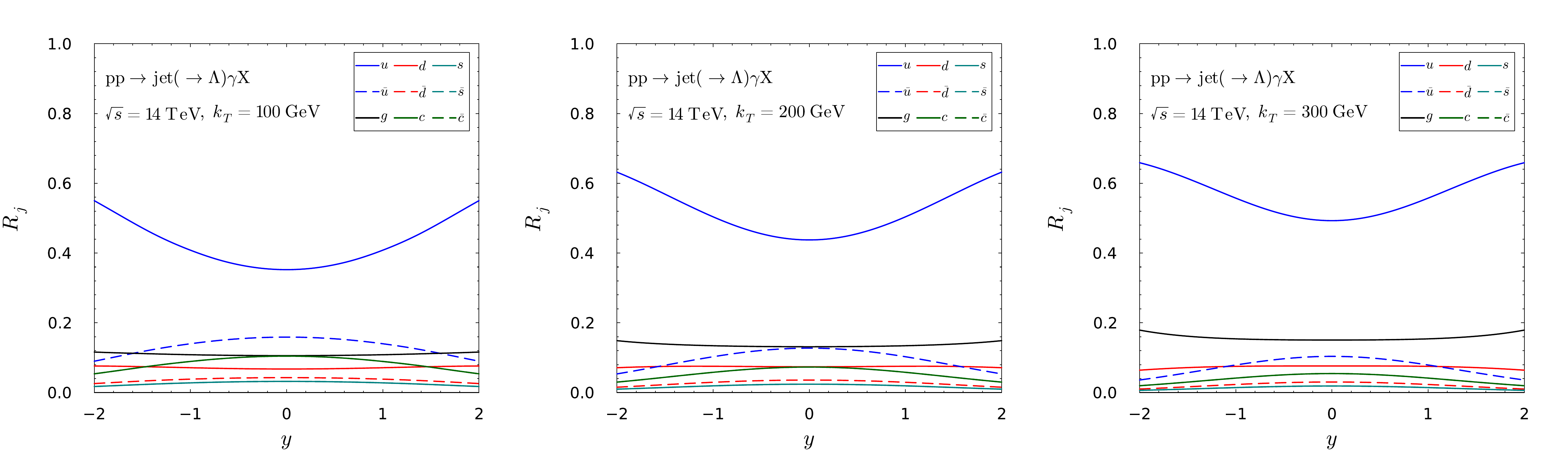}
  \caption{Production ratios of different flavor jets associated with $\gamma$ as functions of $y$ at fixed $k_T$ in $pp$ collisions}
  \label{fig:Rq_pA_gamma}
\end{figure}

Furthermore, for the photon associated jet production, the partonic cross sections are given by
\begin{align}
  &
    \frac{d\hat \sigma_{q+g\to q+\gamma}}{d\hat t} = \pi\alpha_s\alpha_{\rm em}\frac{e_q^2}{3}\left[-\frac{\hat s}{\hat u}-\frac{\hat u}{\hat s}\right],
  \\
  &
    \frac{d\hat \sigma_{q+\bar q\to g+\gamma}}{d\hat t} = \pi\alpha_s\alpha_{\rm em}\frac{8e_q^2}{9}\left[\frac{\hat t}{\hat u}+\frac{\hat u}{\hat t}\right].
\end{align}
Due to the $e_q^2$ factor, the $u$ quark contribution dominates in almost the whole kinematic region. However, as shown in Fig.~\ref{fig:Rq_pA_gamma}, the gluon contribution is flat with increasing rapidity. This is a feature not expected in the single inclusive jet production where the gluon contribution is strongly suppressed in the forward rapidity. Therefore, this process provides with a fresh angle in the global analysis.

\begin{figure}[htb!]
  \centering
  \includegraphics[width=0.9\textwidth]{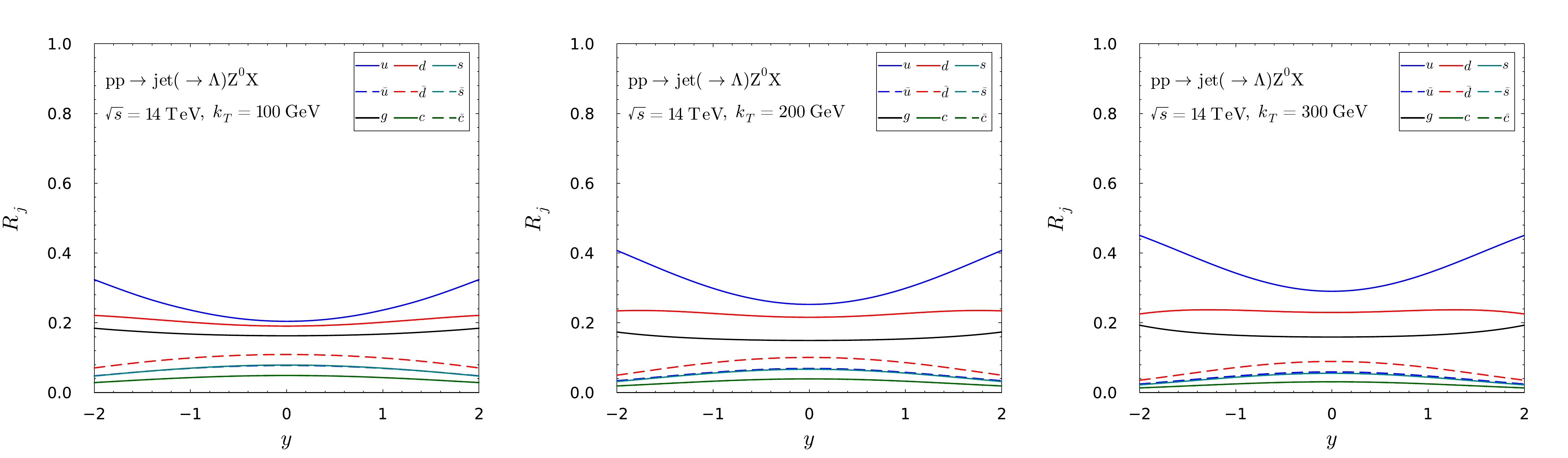}
  \caption{Production ratios of different flavor jets associated with $Z^0$ as functions of rapidity $y$ at fixed $k_T$ at $\sqrt{s}=14~{\rm TeV}$.}
  \label{fig:Rq_pA_Z0}
\end{figure}

On the other hand, for the $Z^0$ associated jet production, the partonic cross sections are given by 
\begin{align}
  &
    \frac{d\hat\sigma_{q+g\to q+Z^0}}{dt} = \pi\alpha_s\alpha_{\rm em}\frac{c_1^q}{3\sin^22\theta_W}\left[-\frac{\hat s}{\hat u}-\frac{\hat u}{\hat s}-\frac{2M_Z^2\hat t}{\hat s\hat u}\right],
  \\
  &
    \frac{d\hat\sigma_{q+\bar q\to g+Z^0}}{d\hat t} =\pi\alpha_s\alpha_{\rm em}\frac{8c_1^q}{9\sin^22\theta_W}\left[\frac{\hat t}{\hat u}+\frac{\hat u}{\hat t}+\frac{2M_Z^2\hat s}{\hat t\hat u}\right].
\end{align}
Here $M_Z$ is the mass of $Z^0$-boson, $\theta_W$ is the Weinberg angle and $c_1^q\equiv (c_V^q)^2+(c_A^q)^2$.
In this case, the contribution from $u$ quark is still important. However, it does not play the dominant role any more. As shown in Fig.~\ref{fig:Rq_pA_Z0}, $u$, $d$ and gluon contributions are, roughly speaking, equally important.

\begin{figure}[htb!]
  \centering
  \includegraphics[width=0.3\textwidth]{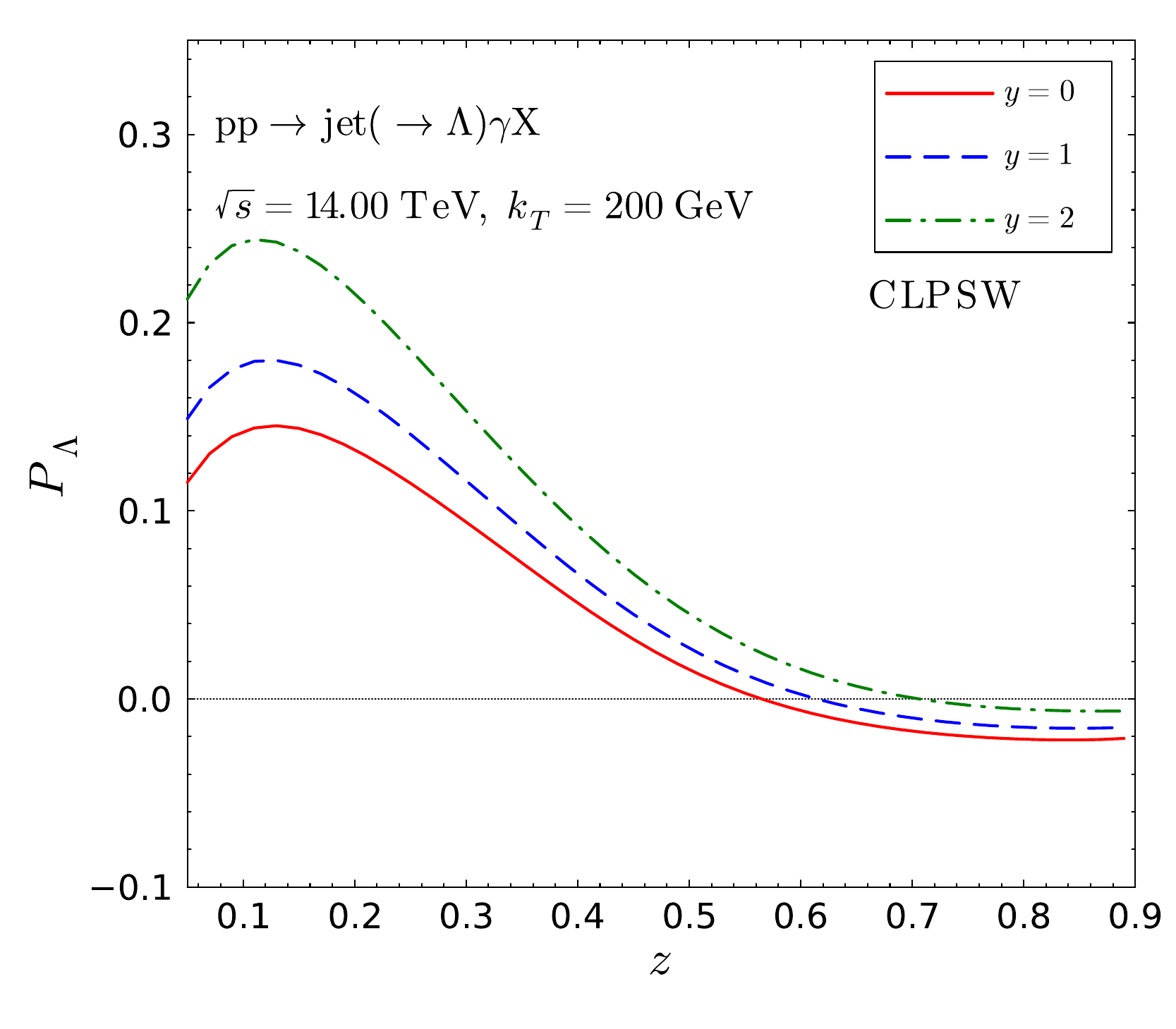}
  \includegraphics[width=0.3\textwidth]{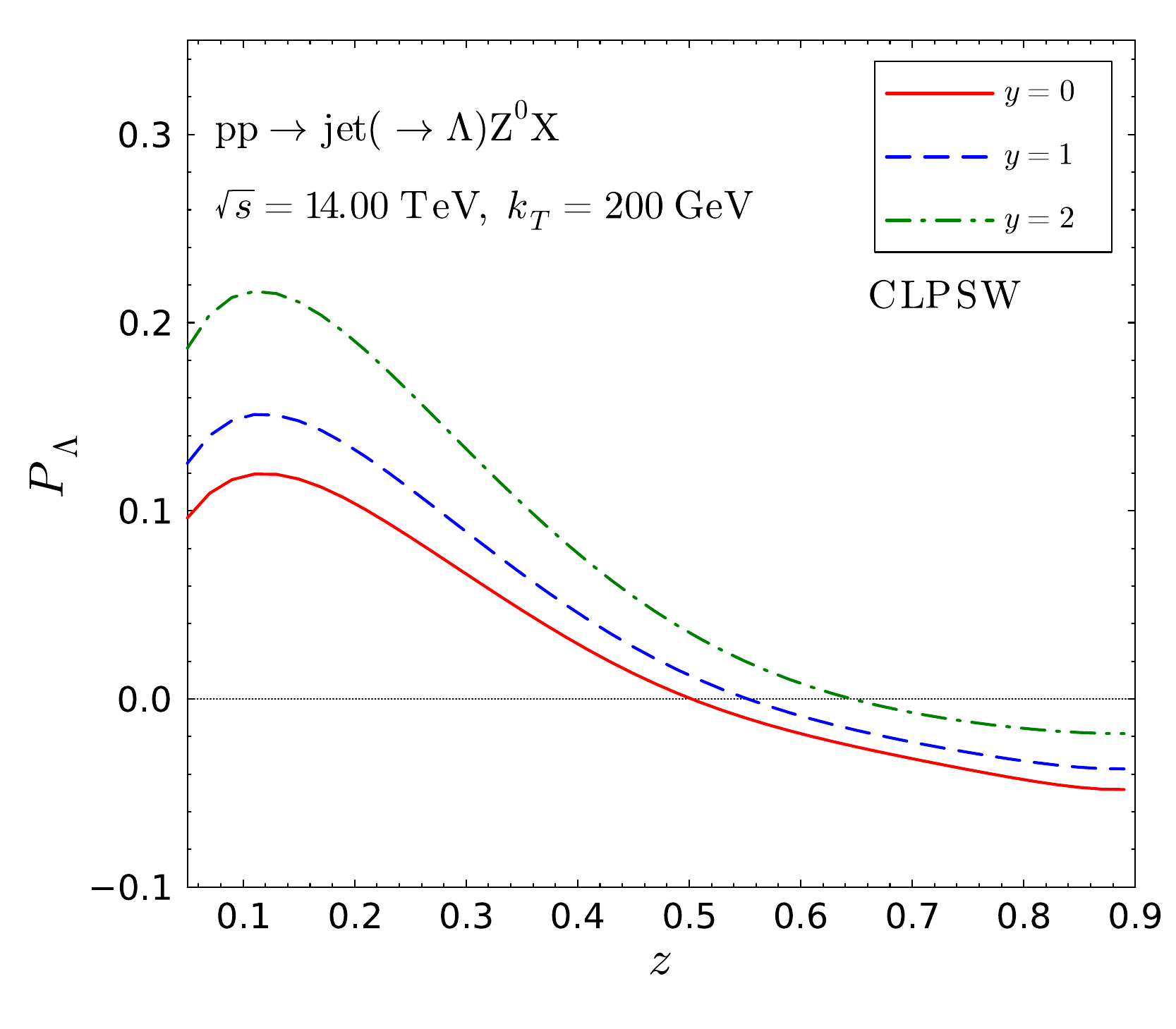}
  \caption{Transverse polarization of $\Lambda$ associated with $\gamma/Z^0$ as functions of $z$ at different $y$ and $k_T$ at $\sqrt{s}=14~{\rm TeV}$.}
  \label{fig:Polarization_pp_gammaZ}
\end{figure}

We present our numerical predictions for $\Lambda$ transverse polarization in the vecor boson associated jet production processes in $pp$ collisions in Fig.~\ref{fig:Polarization_pp_gammaZ}. The left plot is the result for photon-jet production process and the right plot is that for $Z^0$-jet production. Both show that polarization can reach as large as $10\sim 20\%$.

\subsection{Single inclusive jet production in photon-nucleus collisions}

In the relativistic heavy-ion collisions, most of the hard scatterings come from the QCD events where both nuclei break up. However, since the large nucleus is accompanied by an enormous number of quasireal photons, it is also quite plausible that only one of the quasireal photons interact with the other nucleus in the peripheral nucleus-nucleus collisions. In this case, the projectile nucleus (the photon-providing one) does not break up and appears in the final state. The target nucleus (photon-striking one) breaks. In this work, we refer the photon-going direction as the forward direction.

According to the Equivalent Photon Approximation (EPA), we have the following collision
\begin{align}
  \gamma + A \to {\rm jet} (\to \Lambda) + X,
\end{align}
where the photon distribution of the projectile nucleus is given by~\cite{Baltz:2007kq} 
\begin{align}
  x_\gamma f_\gamma(x_\gamma) = \frac{2Z^2 \alpha}{\pi} \left[ \zeta K_0(\zeta)K_1(\zeta) - \frac{\zeta^2}{2} (K^2_1(\zeta) - K^2_0(\zeta)) \right].
\end{align}
Here, $\zeta = x_\gamma M_{{p}} (R_{\rm{A}} + R_{\rm{B}})$ with $x_\gamma$ being the light-cone momentum fraction carried by the quasireal photon. $Z$ is the nuclear charge number, $M_{{p}}$ is the mass of the proton. $R_{\rm{A}}$ and $R_{\rm B}$ are the nucleus radiuses of the projectile and target nuclei. This process resembles SIDIS quite a lot. However, the photon in this process is almost onshell which is different from the case in SIDIS.

The partonic hard scatterings at LO are $\gamma q \to gq$ and $\gamma g \to q\bar q$ with partonic cross sections given by
\begin{align}
  &
    \frac{d\hat \sigma_{\gamma + q\to g + q}}{d\hat t} = \pi\alpha_s\alpha_{\rm em}\frac{8e_q^2}{3}\left[-\frac{\hat s}{\hat u}-\frac{\hat u}{\hat s}\right],
  \\
  &
    \frac{d\hat \sigma_{\gamma + g\to q + \bar q}}{d\hat t} = \pi\alpha_s\alpha_{\rm em}e_q^2\left[\frac{\hat t}{\hat u}+\frac{\hat u}{\hat t}\right].
\end{align}

\begin{figure}[htb!]
  \centering
  \includegraphics[width=0.9\textwidth]{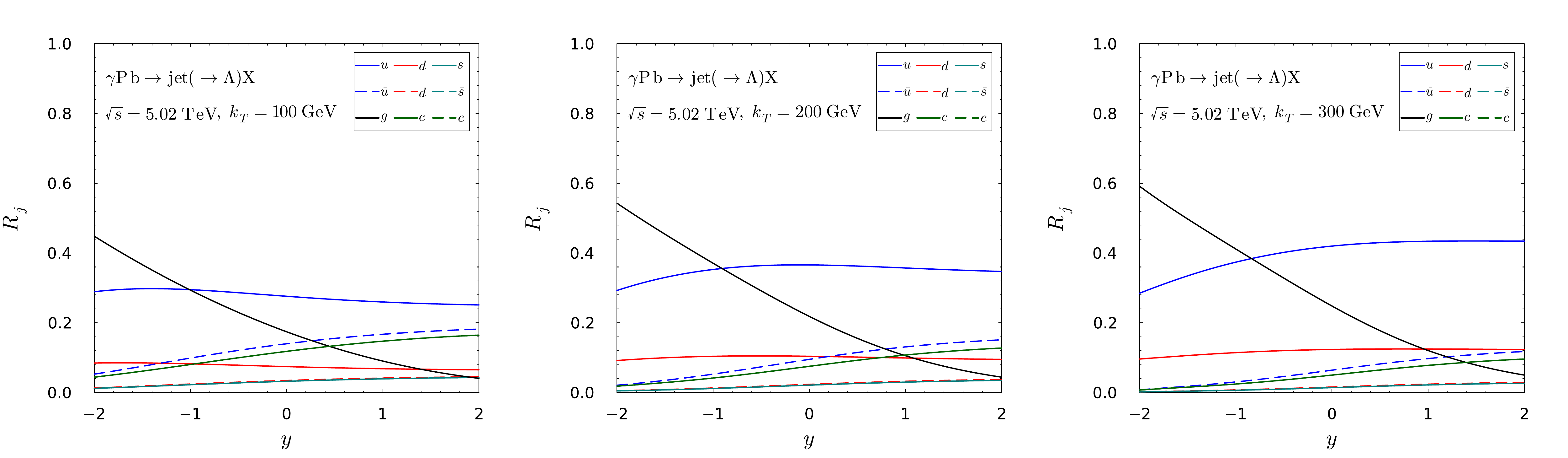}
  \caption{Production ratios for different flavor jets as functions of $y$ at fixed $k_T$ in photon-nucleus collisions. Here and in the following figures the photon distributions inside the lead nuclei is given in~\cite{Baltz:2007kq}.}
  \label{fig:Rq_pA_UPC}
\end{figure}

As shown in Fig.~\ref{fig:Rq_pA_UPC}, the gluon contribution is very important in the backward rapidity. However, it becomes negligible in the forward rapidity. Therefore, this process also offers the optimal platform to investigate the gluon FFs.

\begin{figure}[htb!]
  \centering
    \includegraphics[width=0.3\textwidth]{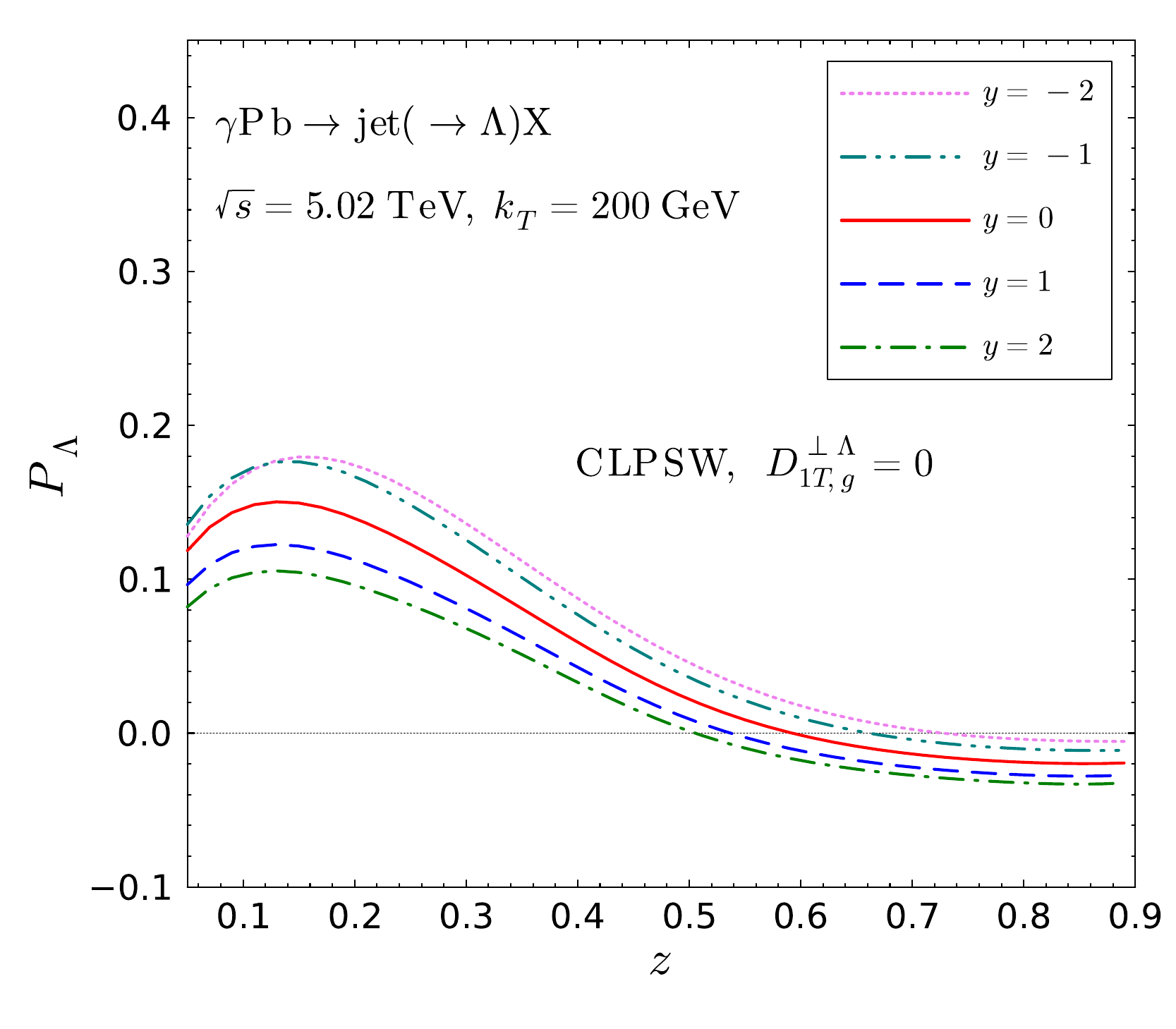}
     \includegraphics[width=0.3\textwidth]{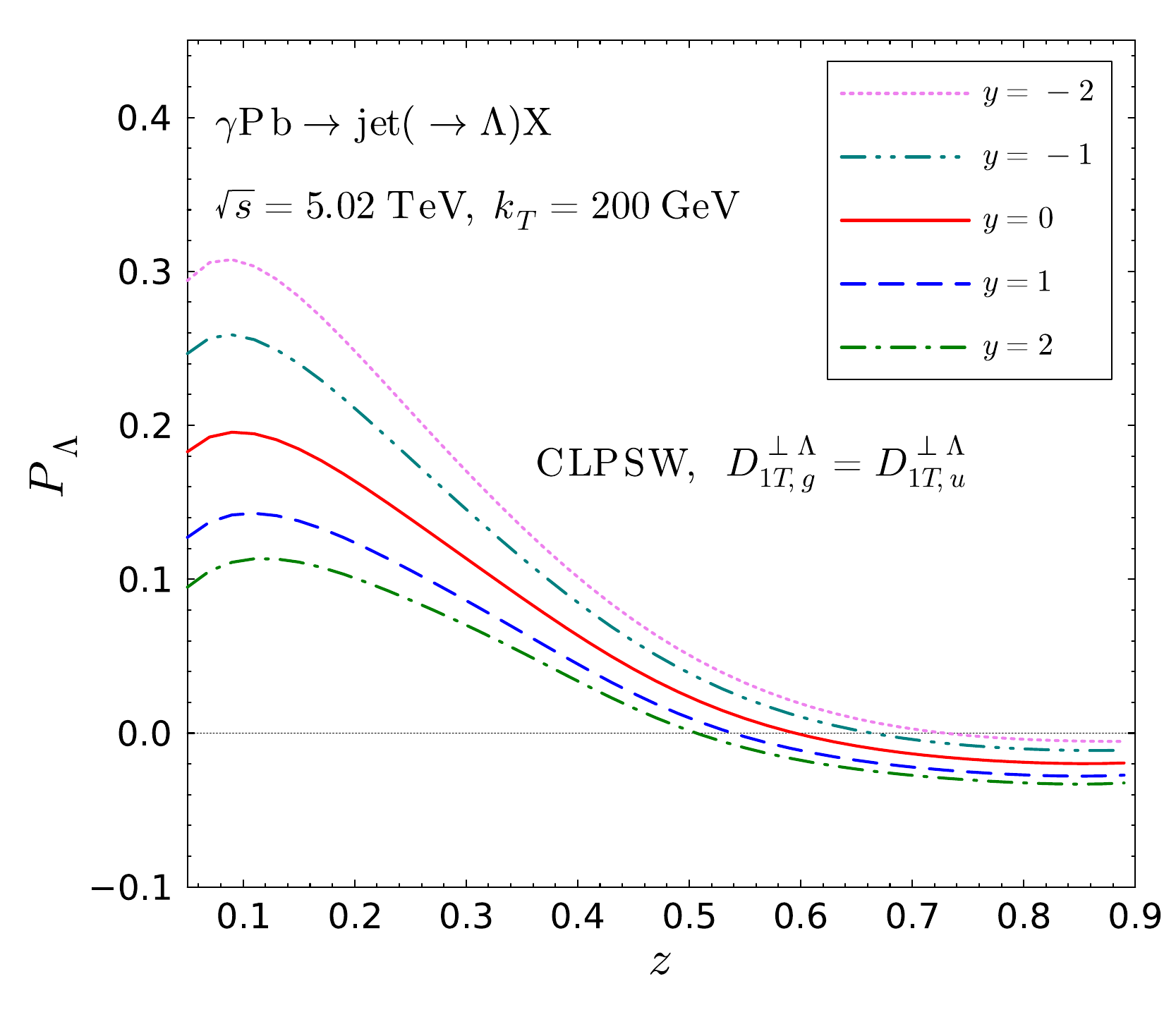}
  \caption{Transverse polarization of $\Lambda$ as functions of $z$ at different $y$ with $k_T=200~{\rm GeV}$ in photon-nucleus collisions.}
  \label{fig:Polarization_AA_UPC}
\end{figure}

We show our numerical predictions for Lambda transverse polarization in photon-nucleus collisions in Fig.~\ref{fig:Polarization_AA_UPC}. The results without gluon contribution are shown in the left panel, and those with gluon contribution are shown in the right panel. As expected, the difference between these two becomes sizable in the backward rapidity, registering the potential to probe the gluon FFs. 

\section{Summary}

The flavor dependence of the polarized fragmentation function $D_{1T}^\perp$ of $\Lambda$ remains ambiguous. While the dispute of isospin symmetry is expected to be ultimately solved \cite{Chen:2021zrr, DAlesio:2023ozw} by the future EIC experiment, the study of polarized gluon FF has to rely on other processes. Recent progress of {\it hadron within jet} \cite{Yuan:2007nd, Bain:2016rrv, Kang:2017glf, Kang:2020xyq, Kang:2023elg} made it possible to investigate TMD FFs in hadronic collisions. In this paper, we demonstrate that the current hadron colliders which are still harvesting data have a great potential in constraining the flavor dependence and extracting the gluon polarized FF. 

In this work, we perform a comprehensive study for the transverse polarization of Lambda hyperons produced in the single inclusive jet production and $\gamma/Z^0$-boson associated jet production processes in $pp$, $p\bar p$, and $pA$ collisions. By comparing the Lambda transverse polarization in different kinematic regions and in different processes, we can gain information on the flavor dependence. Particularly, for the single inclusive jet production in $pp$ collisions, the dominant contribution comes from gluon in the central rapidity, while the dominant contribution comes from $u$ quark in the forward rapidity. Therefore, by comparing the Lambda polarizations in forward and central rapidities, we can separate gluon FF from the $u$ quark FF. Similarly, in $p\bar p$ collisions, the dominant contribution arises from gluon as well in the central rapidity. However, $u$ quark dominates in the forward rapidity, while $\bar u$ quark dominates in the backward rapidity. Such a forward-backward asymmetry allows us to separate $\bar u$ quark FF from $u$ quark FF. A forward-backward asymmetry also appears in $pA$ collisions, which allows us to test the isospin symmetry, i.e., the difference between $u$ and $d$ FFs. Furthermore, the flavor component of the $\gamma/Z^0$-boson associated jet production is different from that of the single inclusive jet production. Therefore, it casts more light on the flavor dependence of the polarized FF.

As the complementary approach, we also study the Lambda transverse polarization in the photon-nucleus collisions. In this case, the gluon contribution varies significantly with the rapidity, which makes this process a novel probe to explore the polarized gluon FF.

\begin{acknowledgments}
  
  This work is supported in part by the National Natural Science Foundation of China (approval number 12375075, 11505080, 12005122), the Taishan fellowship of Shandong Province for junior scientists, the Shandong Province Natural Science Foundation under grant No. 2023HWYQ-011, No. ZR2018JL006, and No. ZR2020QA082, and the Youth Innovation Team Program of Higher Education Institutions in Shandong Province (Grant No. 2023KJ126).
  
\end{acknowledgments}

\appendix
\section{Production ratios and polarization at LHC}

In this appendix, we present our predictions for the single inclusive jet production in $pp$ collisions at LHC. 

\begin{figure}[htb!]
  \centering
  \includegraphics[width=0.9\textwidth]{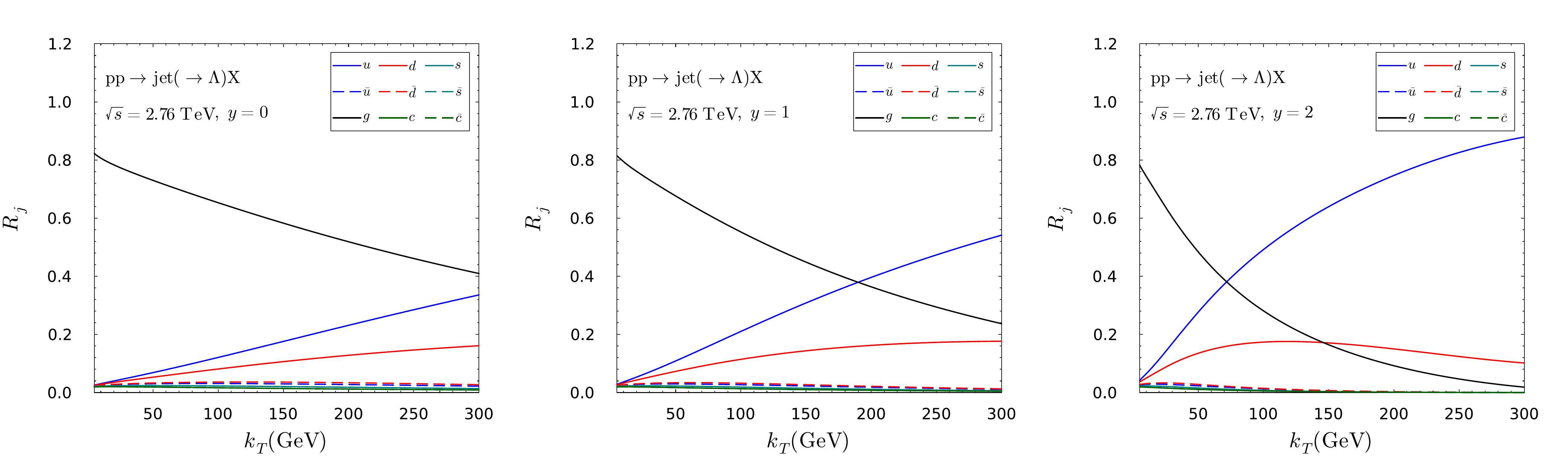}\\
  \includegraphics[width=0.9\textwidth]{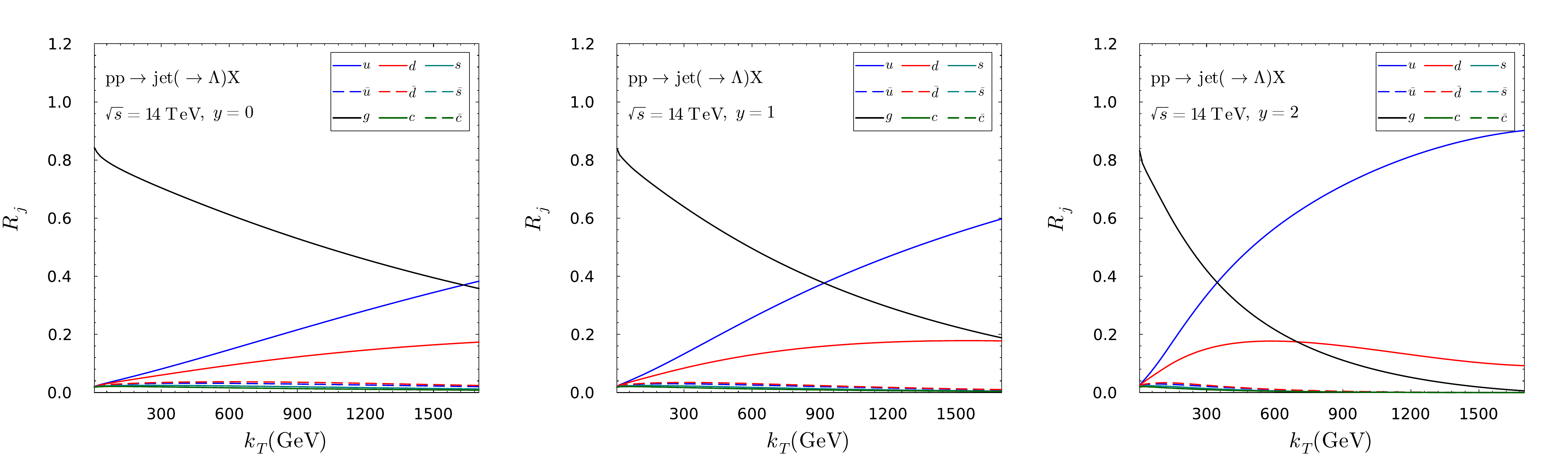}
  \caption{Production ratios of different flavor jets as functions of $k_T$ at fixed rapidity $y$ in $pp$ collisions at LHC.}
  \label{fig:Rq_pp_y0_LHC}
\end{figure}

\begin{figure}[htb!]
  \centering
  \includegraphics[width=0.9\textwidth]{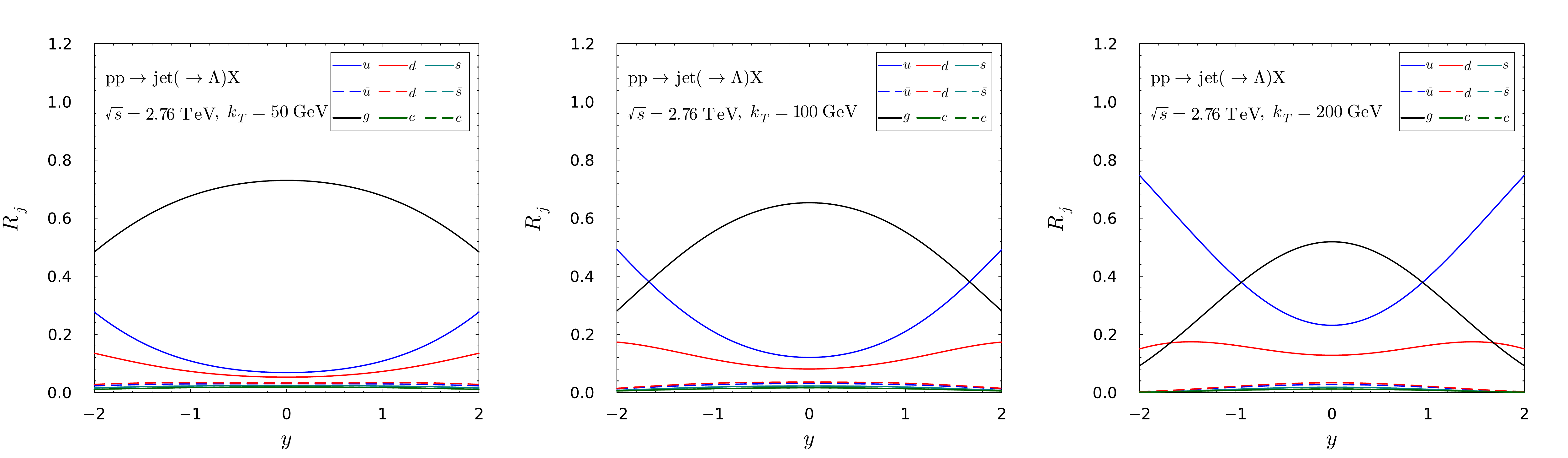}\\
  \includegraphics[width=0.9\textwidth]{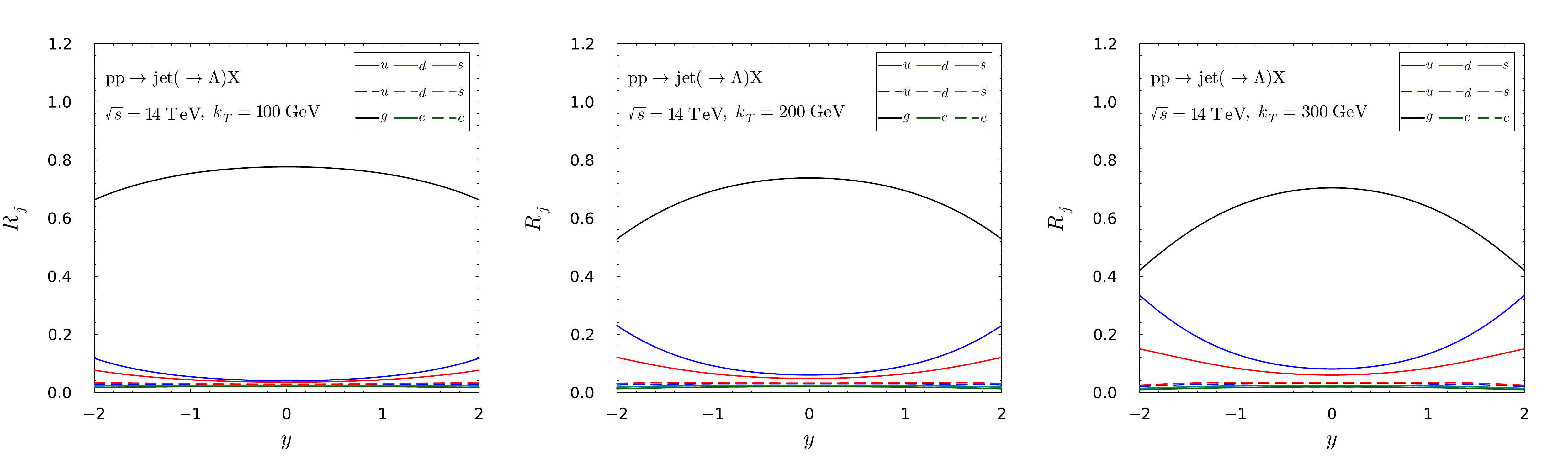}
  \caption{Production ratios of different flavor jets as functions of $y$ at fixed rapidity $k_T$ in $pp$ collisions at LHC.}
  \label{fig:Rq_pp_pt0_LHC}
\end{figure}

We first present the production ratios in Figs.~\ref{fig:Rq_pp_y0_LHC} and \ref{fig:Rq_pp_pt0_LHC} as functions of $k_T$ at fixed rapidity $y$, and as functions of $y$ at fixed $k_T$ with $\sqrt{s}=2.76~{\rm TeV},~14~{\rm TeV}$ at the LHC. 

\begin{figure}[htb!]
  \centering
  \includegraphics[width=0.9\textwidth]{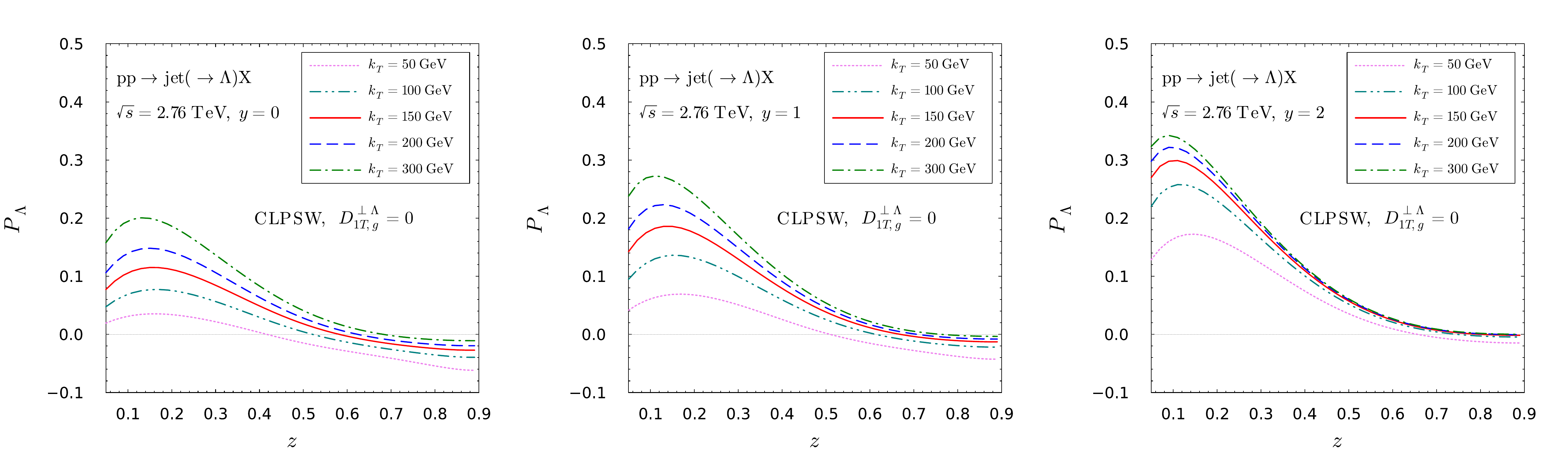}\\
  \includegraphics[width=0.9\textwidth]{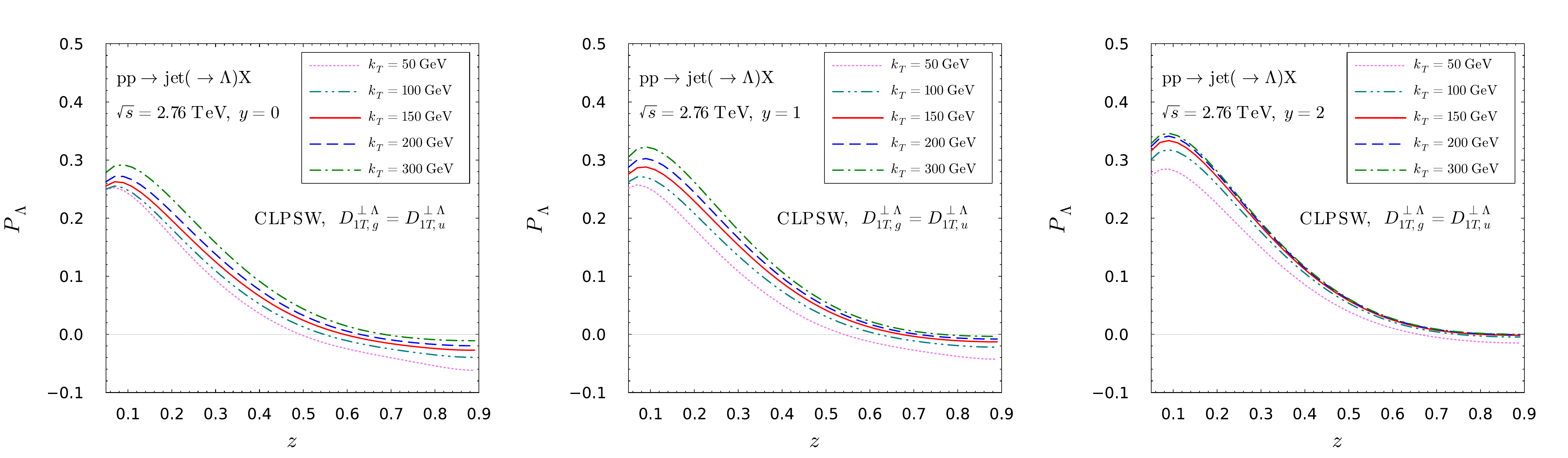}
  \caption{Transverse polarization of $\Lambda$ as functions of $z$ at different $y$ and $k_T$ in $pp$ collisions with $\sqrt{s}=2.76~{\rm TeV}$ at LHC. The upper panel shows the numerical results without gluon contributions, while the lower panel shows the results with gluon contributions.}
  \label{fig:Polarization_pp_2.76TeV}
\end{figure}

\begin{figure}[htb!]
  \centering
  \includegraphics[width=0.9\textwidth]{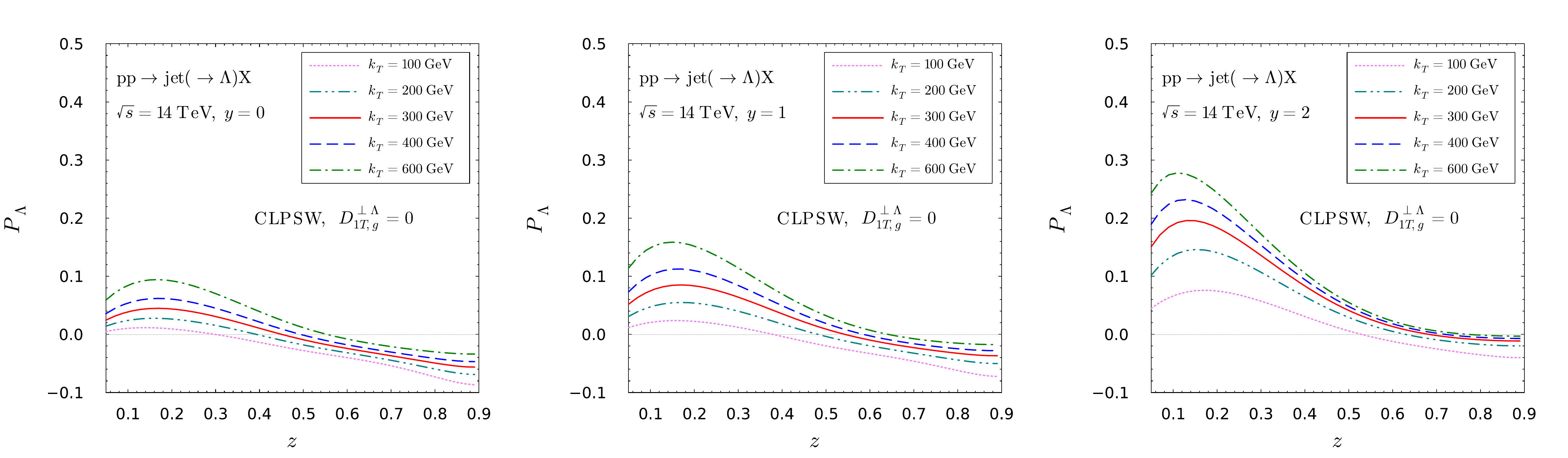}\\
  \includegraphics[width=0.9\textwidth]{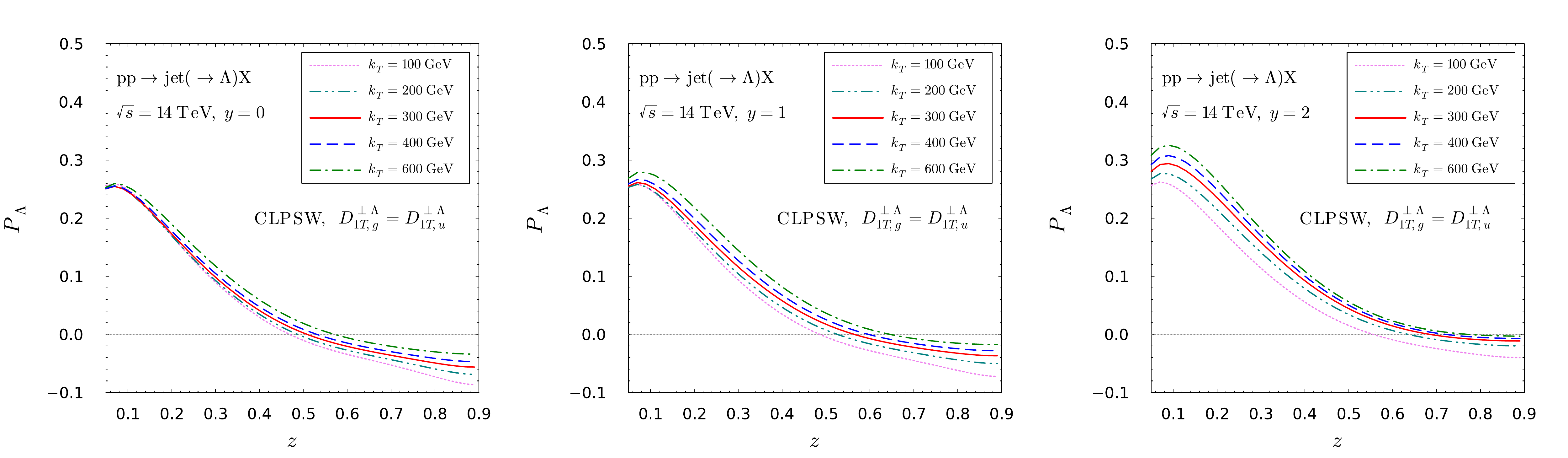}
  \caption{Transverse polarization of $\Lambda$ as functions of $z$ at different $y$ and $k_T$ in $pp$ collisions with $\sqrt{s}=14~{\rm TeV}$ at LHC. The upper panel shows the numerical results without gluon contributions, while the lower panel shows the results with gluon contributions.}
  \label{fig:Polarization_pp_14TeV}
\end{figure}

Then we present our numerical results for the Lambda transverse polarization at LHC in Fig. \ref{fig:Polarization_pp_2.76TeV} and \ref{fig:Polarization_pp_14TeV}.


\clearpage

\section{Next-to-leading order correction}

The leading-order calculation contains large uncertainties and is not sufficient for precision studies. However, the existing experimental data is still far from being accurate enough to constrain the polarized fragmentation function. Therefore, in this paper, we only present a general guideline on how to disentangle the flavor dependence of $D_{1T}^\perp$ in current hadron colliders, and leave a more accurate next-to-leading order calculation for a future study when more experimental data with a higher precision is available. In this section, we simply demonstrate that the qualitative features obtained based on the leading order calculation is not affected by higher order corrections. 

\begin{figure}[htb]\centering
\includegraphics[width=0.3\textwidth]{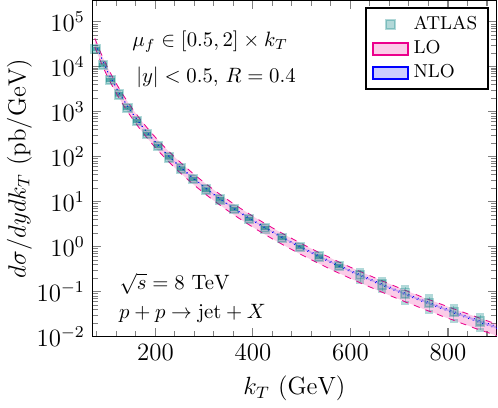}
\caption{Leading and next-to-leading order results for the single inclusive jet production in unpolarized pp collisions compared with experimental data from the ATLAS collaboration \cite{ATLAS:2017kux} at the LHC.}
\label{fig:cs}
\end{figure}

In Fig.~\ref{fig:cs}, we first show the single inclusive jet production cross section at leading and next-to-leading orders in unpolarized pp collisions at $\sqrt{s} = 8$ TeV compared with the experimental data from the ATLAS collaboration \cite{ATLAS:2017kux} at the LHC. We have varied the factorization/renormalization scales $\mu_f = \mu_r \in [0.5, 2] k_T$ with $k_T$ being the jet transverse momentum to estimate the theoretical uncertainty. As expected, the scale dependence at the next-to-leading order is much smaller than that at the leading order. However, by properly choosing scales (for instance $\mu_f = \mu_r = k_T$), we can eliminate large logarithms in the next-to-leading order hard factor, and therefore minimize higher-order corrections. This way, even the leading order calculation can also describe the ATLAS data well. 

\begin{figure}[htb]\centering
\includegraphics[width=0.6\textwidth]{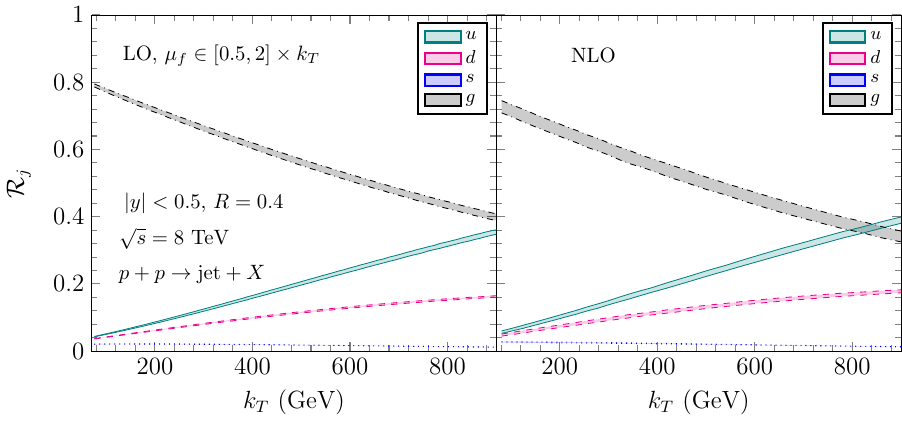}
\caption{Flavor components of jets as a function of transverse momentum.}
\label{fig:ratio}
\end{figure}

Furthermore, we also present the leading and next-to-leading order jet flavor components as a function of jet transverse momentum in Fig.~\ref{fig:ratio}. We only present the relative contribution from $u$, $d$, $s$, and gluon in Fig.~\ref{fig:ratio}. Contributions from the other partons are not shown explicitly since they are not of interest here. Clearly, the qualitative conclusion based on the leading order calculation remains unaltered at the next-to-leading order. Gluon jet still dominates at low $k_T$ and $u$ quark contribution significantly rises with increasing $k_T$.

To summarize, the leading order calculation offers an intuitive and correct physical picture on the flavor component. The higher-order contribution is expected to provide minor correction to the flavor component and will not affect our conclusion. 

\section{Collins-Soper Evolution of $D_{1T}^\perp (z, p_T,\mu_f)$}

In the hybrid factorization scheme, the jet fragmentation assumes the TMD factorization. Both the unpolarized jet fragmentation function $D_1 (z,p_T, \mu_f)$ and the polarized fragmentation function $D_{1T}^\perp (z, p_T, \mu_f)$ satisfy the Collins-Soper evolution equation. Here, we have already set all scales to be the same for simplicity. In the end, we arrive at
\begin{align}
&
D_{1} (z, p_T, \mu_f) = \int \frac{d^2 \bm{b}}{(2\pi)^2} e^{-i \bm{p}_T \cdot \bm{b}/z} \exp \left[-S_{\rm pert} (\mu_f, \mu_b) - S_{\rm np} (b) \right] D_{1} (z, \mu_b) ,
\\
&
D_{1T}^\perp (z, p_T, \mu_f) = \frac{M_\Lambda^2}{zp_T} \int \frac{d^2 \bm{b}}{(2\pi)^2} b e^{-i\bm{p}_T \cdot \bm{b}/z}  \exp \left[-S_{\rm pert} (\mu_f, \mu_b) - S_{\rm np}^T (b) \right] D_{1T}^{\perp (1)} (z, \mu_b).
\end{align}
Here we have employed the $b^*$-prescription to separate the perturbative and the non-perturbative regions, so that the scale $\mu_b = 2e^{-\gamma_E}/b^* \ge 2e^{-\gamma_E}/b_{\max} \gg \Lambda_{\rm QCD}$ with $b_{\max} \sim 1$ GeV$^{-1}$. 

The exact expressions for perturbative and non-perturbative Sudakov factors are taken from Ref.~\cite{Gamberg:2021iat}. The perturbative Sudakov factor $S_{\rm pert} (\mu_f,\mu_b)$ remains the same for the polarized and unpolarized fragmentation functions. $S_{\rm np}$ and $S_{\rm np}^T$ are non-perturbative Sudakov factors that can only be extracted from experimental data. We note here that the non-perturbative Sudakov factor still contains large uncertainties, which can only be removed through a global analysis to the future experimental data.

$D_1 (z, \mu_b)$ is the collinear fragmentation function and therefore follows the DGLAP evolution equation. On the other hand, the QCD evolution of $D_{1T}^{\perp (1)} (z, \mu_b)$ is a bit more involved. First, it is not a closed equation \cite{Kang:2010xv}. The evolution kernel consists of a conventional DGLAP evolution part and also a twist-3 source term. The twist-3 source term is so far unknown. Second, even the conventional DGLAP evolution part also evolves the gluon polarized fragmentation function which is also not well constrained by the Belle experiment. At this stage of research, we have neglected the QCD evolution of the collinear $D_{1T}^{\perp (1)} (z, \mu_b)$ function in the numerical evaluation and simply assume 
\begin{align}
D_{1T}^{\perp (1)} (z, \mu_b) = \int d^2 p_T \frac{p_T^2/z^2}{2M_\Lambda^2} D_{1T}^\perp (z, p_T, \mu_0), \label{eq:c2}
\end{align}
with $\mu_0 = 10$ GeV being the factorization scale where $D_{1T}^\perp (z, p_T, \mu_f = \mu_0)$ has been extracted.

\begin{figure}[htb] \centering
\centering
\includegraphics[width=0.3\textwidth]{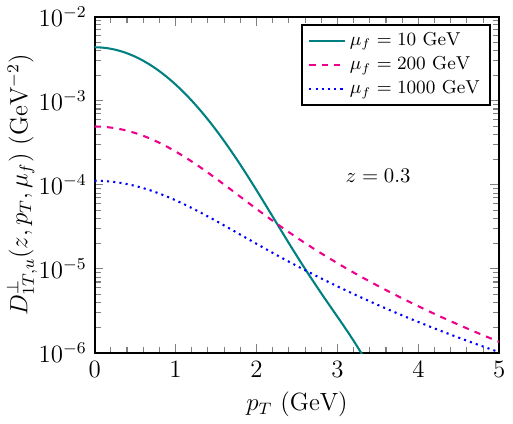}
\caption{The polarized fragmentation function $D_{1T,u}^\perp(z,p_T,Q)$ as a function of $p_T$ at different factorization scales.}
\label{fig:tmdff}
\end{figure}

With the above configuration, we first present $D_{1T}^\perp (z, p_T, \mu_f)$ as a function of $p_T$ at different factorization scales in Fig.~\ref{fig:tmdff}. Thanks to the Sudakov double and single logarithms, the $p_T$ broadening effect is quite strong at large factorization scale. 

\begin{figure}[htb]\centering
\includegraphics[width=0.6\textwidth]{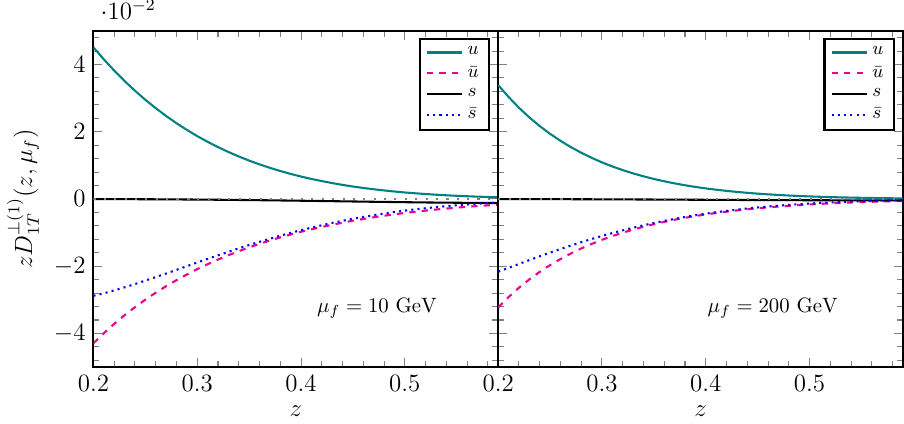}
\includegraphics[width=0.6\textwidth]{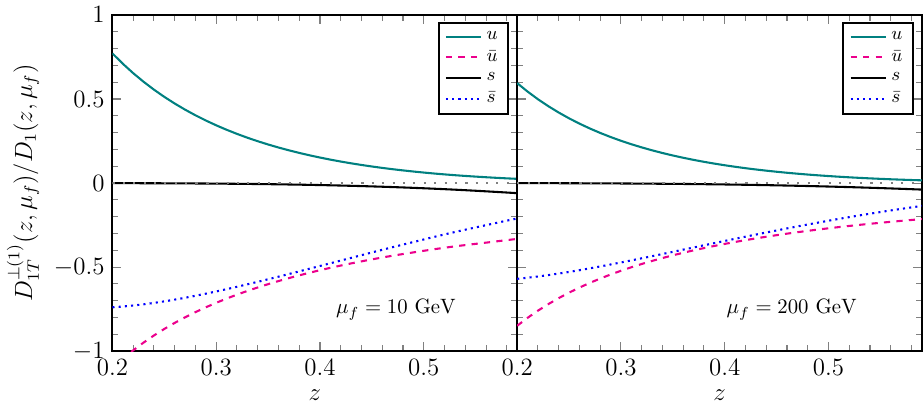}
\caption{The first moments $D_{1T}^{\perp(1)} (z,\mu_f)$ of different flavors as a function of $z$ at different factorization scales, and their ratios to the unpolarized FFs $D_1(z,\mu_f)$.}
\label{fig:collff}
\end{figure}

Since we have integrated over the $p_T$ in this research, it is intriguing to present the first moment of $D_{1T}^{\perp}$ at different factorization scales as well. In contrast with Eq.~(\ref{eq:c2}), $D_{1T}^{\perp (1)} (z,\mu_f)$ is defined as
\begin{align}
D_{1T}^{\perp (1)} (z,\mu_f) = \int_0^{p_{T,{\max}}} d^2 p_T \frac{p_T^2/z^2}{2M_\Lambda^2} D_{1T}^\perp (z, p_T, \mu_f),
\end{align} 
where $p_{T,{\max}} = 5$ GeV. The numerical results are presented in Fig.~\ref{fig:collff}. The QCD evolution affects the absolute magnitude of $D_{1T}^{\perp(1)}(z,\mu_f)/D_1(z,\mu_f)$ only by a little bit. 

\end{document}